\documentclass[12pt]{elsarticle}
\usepackage{epsfig,graphics,amssymb,amsmath,graphicx,indentfirst,subfig,float,appendix,multirow,pifont,color,soul,esint}
\usepackage{threeparttable}
\usepackage{graphicx}
\usepackage[dvipsnames]{xcolor}
\usepackage{lineno}
\providecommand{\keywords}[1]
{
  \small	
  \textbf{\textit{Keywords---}} #1
}

\begin{document}
\begin{frontmatter}

\title{Analysis of the suitability of an effective viscosity to represent interactions between red blood cells}

\author[cornell]{Grant Rydquist}
\author[cornell]{Mahdi Esmaily$^*$}
\address[cornell]{Sibley School of Mechanical and Aerospace Engineering, Cornell University, Ithaca, NY 14850, USA}

\date{\today} 
\begin{abstract}

Many methods to computationally predict red blood cell damage have been introduced, and among these are Lagrangian methods which track the cells along their pathlines. Such methods typically do not explicitly include cell-cell interactions. Due to the high volume fraction of red blood cells in blood, these interactions could impact cell mechanics and thus, the amount of damage caused by the flow. To investigate this question, cell-resolved simulations of red blood cells in shear flow were performed for multiple interacting cells, as well as for single cells in unbounded flow at an effective viscosity. Simulations run without adjusting the bulk viscosity produced larger errors unilaterally and were not considered further for comparison. We show that a periodic box containing at least 8 cells and a spherical harmonic of degree larger than 10 are necessary to produce converged higher-order statistics. The maximum difference between the single-cell and multiple-cell cases in terms of peak strain was 3.7\%. To achieve this agreement, one must use the whole blood viscosity and average over multiple cell orientations when adopting a single-cell simulation approach. There were some differences between the two models in terms of average strain (maximum difference of 13\%). However, given the accuracy of the single-cell approach in predicting the maximum strain, which is useful in hemolysis prediction, and its computational cost that is orders of magnitude less than the multiple-cell approach, one may use it as an affordable cell-resolved approach for hemolysis prediction. 

% END CONTENT ABS------------------------------------------
%\noindent

$^*$ Corresponding author: me399@cornell.edu

\textbf{Acknowledgements}: Research reported in this publication was supported by the National Heart, Lung, and Blood Institute of the National Institutes of Health under award number R01HL089456-10.
\end{abstract}
\end{frontmatter}
\newpage

% %%%%%%%%%%%%%%%%%%%%%%%%%%%%%%%%%%%%%%%%%%%%%%%%%%%%%%%%%%
% %%%%%%%%%%%%%%%%%%%%%%%%%%%%%%%%%%%%%%%%%%%%%%%%%%%%%%%%%%
% BODY OF THE DOCUMENT
% %%%%%%%%%%%%%%%%%%%%%%%%%%%%%%%%%%%%%%%%%%%%%%%%%%%%%%%%%%
% %%%%%%%%%%%%%%%%%%%%%%%%%%%%%%%%%%%%%%%%%%%%%%%%%%%%%%%%%%

% --------------------
\section{Introduction} \label{sec:introduction}

There have been several models introduced to investigate the effects of a flow field on red blood cells (RBCs), particularly concerning hemolysis \cite{yu_review_2017, giersiepen_estimation_1990, sharp_scaling_1998, arora_tensor-based_2004, arvand_validated_2005, ezzeldin_strain-based_2015, sohrabi_cellular_2017, nikfar_multiscale_2020}. Typically, when attempting to characterize hemolysis, these methods take a continuum view of the RBC phase, in which the whole blood is modeled as a single fluid{. A common framework for evaluating hemolysis under these assumptions is an Eulerian one in which stress and residence time are evaluated at fluid elements throughout the entire flow to obtain a picture of how the flow might damage the RBCs \cite{garon_fast_2004, zhang_computational_2006}.} Another common method is to follow Lagrangian pathlines through the blood{, which itself is still treated as an Eulerian continuum,} and use properties of the fluid integrated along these pathlines to solve empirical equations representing the amount of damage occurring {\cite{grigioni_power-law_2004, goubergrits_numerical_2004}}. However, recently some Lagrangian models have been introduced that explicitly track RBCs as they move through the flow and resolve their behavior {at a microscale level} \cite{arora_tensor-based_2004, ezzeldin_strain-based_2015, sohrabi_cellular_2017, nikfar_multiscale_2020}. These types of methods can be useful in that they can provide a more complete picture of a RBC's response in comparison to those where the RBCs are not tracked explicitly. The fluid-based types of models are typically based on empirical relationships which are not necessarily suitable for arbitrary flows, whereas the cell-resolved models attempt to replicate the behavior of the RBCs directly. For example, several fluid-based methods to determine hemolysis in turbulent flows have been introduced \cite{tamagawa_prediction_1996, ozturk_approach_2016, goubergrits_turbulence_2016}. However, there is disagreement on the effectiveness of these methods, or if they are incorporating the correct physical phenomena.

\begin{figure}[H]
    \centering
        \includegraphics[scale=.31]{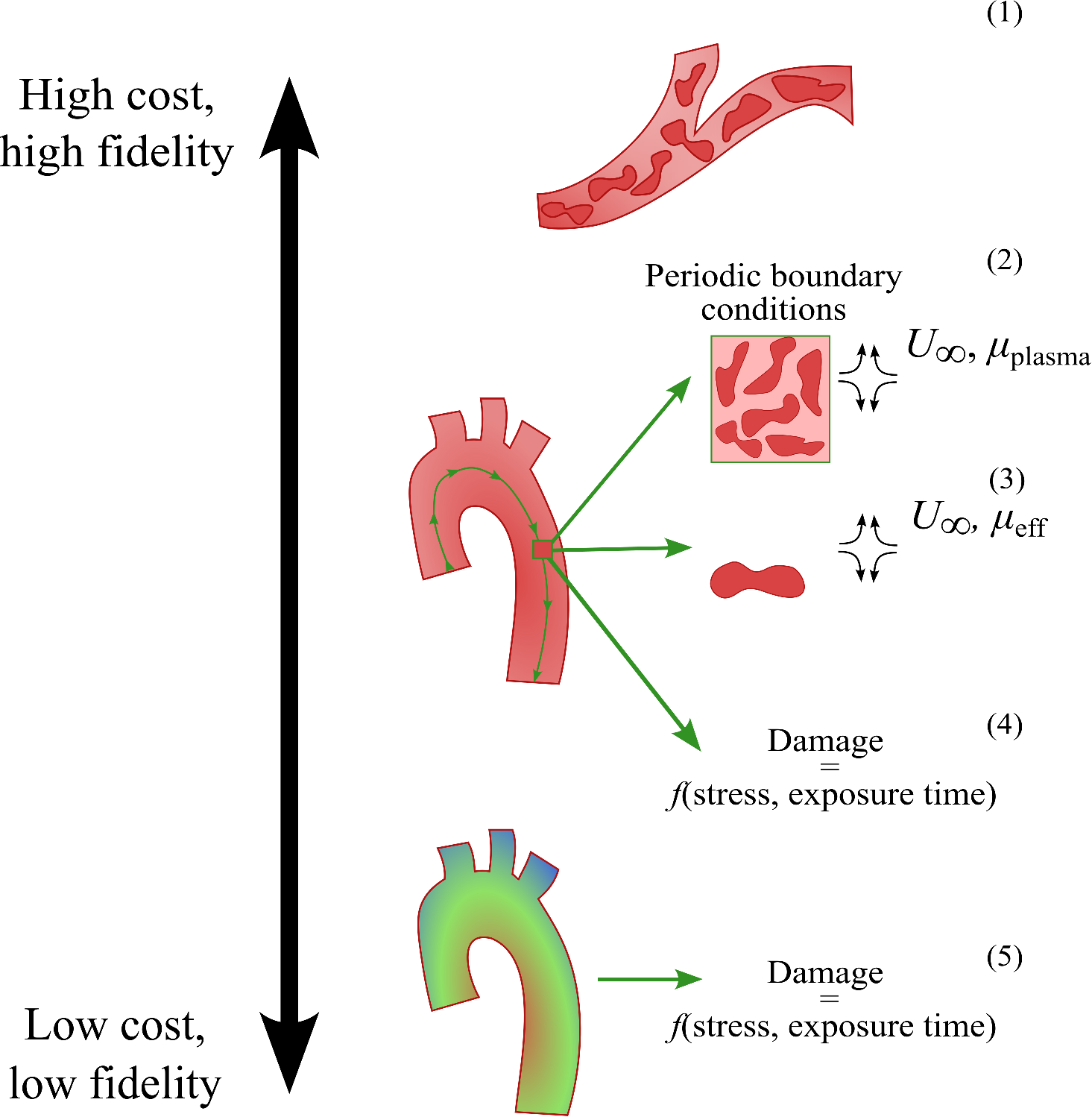}
    \caption{Different approaches for modeling the response of RBCs to a fluid flow. The most computationally expensive methods simulate the coupling between the RBCs and the geometry (1), with such methods typically only being feasible for microvasculature. Lower fidelity methods do not resolve these interactions exactly, but instead track pathlines through the geometry of interest. The pathlines can represent a periodic box of cells, which preserves cell-cell interactions (2) or a single cell in which an effective viscosity must be utilized (3), or parameters such as stress and strain along the pathline can be used as an input to a function representing the amount of damage occurring (4). Finally, the damage occurring can be evaluated in a fully Eulerian approach, where pathlines are not tracked and cell damage is calculated at specific locations using empirical relationships (5).}
    \label{fig:cost_fidelity}
\end{figure}

The greater fidelity simulations generally come at the expense of higher computational cost. The above simulation strategies are represented schematically in Fig. \ref{fig:cost_fidelity}, ranging from highest cost and highest fidelity at the top, to lower cost, lower fidelity at the bottom. In the highest resolution case, both the RBCs and surrounding geometry are represented exactly and solved together \cite{balogh_direct_2017}. Due to the large number of cells required for this approach, it is typically only suitable for small vessels, although some studies have been performed on larger geometries using high-performance computing \cite{lu_scalable_2019, peters_multiscale_2010}. Even so, the vessels simulated were smaller than what might be of interest for hemolysis. Below this, the approach which is explored in the current paper is to resolve both the cells and the vessels, but to only incorporate the local flow effects on the RBCs and model the effect of the RBCs on the fluid through a rheological model. Such an approach has been utilized somewhat \cite{ezzeldin_strain-based_2015, sohrabi_cellular_2017, nikfar_multiscale_2020, rydquist_cell-resolved_2022}, but is relatively unexplored. In this approach, one would solve a fluid-structure interaction problem that involves multiple interacting cells in an unbounded flow at zero Reynolds number. Hypothetically, the cell-cell interactions can be incorporated directly through the use of periodic boundary conditions. To simplify computations further, one may also model the cell-cell interactions by adjusting the fluid bulk viscosity that results from the stress in the suspension. All of the cell-resolved approaches thus far implicitly make this assumption that the cell-cell interactions can be properly incorporated through the bulk viscosity. The resulting approach thus still solves a fluid-structure interaction problem but with far fewer degrees of freedom than the periodic approach. Below these single-cell simulations with an effective viscosity is the Lagrangian approach in which pathlines are tracked through the fluid and, instead of resolving the cells directly, empirical equations to calculate the risk of hemolysis are used. Finally, below this is the Euler-Euler approach, in which specific regions of the flow are analyzed using empirical equations that use stress and exposure time as inputs. {Often, these methods are obtained via a linearization of power law damage functions, which can lead to some mathematical errors \cite{faghih_modeling_2019}. For example, they may be applicable only in uniaxial flows and may be mathematically inconsistent with the power laws they are originally derived from \cite{faghih_modeling_2019}.} The current work is focused on a comparison of approaches (2) and (3) in Fig. \ref{fig:cost_fidelity}. Specifically, the purpose of the current work is to explore the importance of cell-cell interactions in the cell-resolved Lagrangian approaches, and the extent to which they impact some key parameters in the calculation of RBC damage.
By adopting a cell-resolved approach, one may circumvent the difficulties associated with the non-generality of the fluid-based methods. That is, the cells are agnostic to the actual character of the larger flow as long as the flow in their direct vicinity is resolved properly. For cell-resolved approaches, several factors affect the flow behavior in the vicinity of RBCs that ultimately determine the accuracy of these computations. Among those, a factor of prime importance is the interactions between cells. RBCs are densely packed in blood, reaching volume fractions of approximately $45-50\%$. As a result, interactions between cells could be essential in determining their overall behavior.

{The RBCs are important drivers of the rheology of blood, and their impact on this rheology has been studied experimentally, theoretically, and computationally. Blood is shear-thinning, and the seminal paper by Chien \cite{chien_shear_1970} was among the first to explain this behavior from a theoretical and experimental perspective. At lower shear rates, Chien posits that the increased effective viscosity is primarily caused by aggregation of the cells. At these low shear rates, the cells form rod-like aggregates known as rouleaux. Chien compares the viscosity of RBCs suspended in plasma to RBCs suspended in an albumin-Ringer solution of the same viscosity in which the cells do not form rouleaux, showing a large increase in viscosity in the experiments where the cells do form rouleaux. At higher shear rates, the rouleaux break up, and the viscosities of these two solutions collapse as a result, suggesting that aggregation becomes unimportant to the rheology of blood at high shear rates. At these shear rates, the shear-thinning behavior of the cells is primarily attributed to the cell deformation. Chien also measures the viscosity of a solution of RBCs in an albumin solution that have been hardened such that they do not deform or aggregate. Across all shear rates, the viscosity of this solution is fairly constant relative to the other solutions, suggesting that deformation of the RBCs plays a role in the shear-thinning behavior of blood, particularly at high shear rate. As RBC simulation has become more tenable, these experimental results have been corroborated computationally \cite{fedosov_predicting_2011}.}

Incorporating {the interactions between cells} is a computational hurdle to cell-resolved methods: these interactions decay slowly, and, naively, an infeasibly large number of cells would need to be included in the simulations to accurately resolve these effects on just a single cell. An experimental study by Leverett et al. shows these interactions may not affect the hemolysis behavior of the cells \cite{leverett_red_1972}. However, recent numerical experiments by Porcaro and Saeedipour show that these interactions may affect hemolysis \cite{porcaro_hemolysis_2023}. These simulations were performed in a microfluidic channel with a coarse RBC model, however, whereas the current work focuses on a highly accurate RBC model in unbounded flow, which is more suitable for the context of evaluation of hemolysis in macro-scale flows.

By comparing approaches (2) and (3) in Fig. \ref{fig:cost_fidelity}, this study establishes scenarios under which cell-cell interactions can be neglected in a cell-resolved simulation. The underlying framework for the simulations performed in this work is described in our previous work \cite{rydquist_cell-resolved_2022}, and the methods will only briefly be revisited when necessary. In short, we solve for the velocity on the surface of the RBC using the boundary integral method (BIM). {It should be noted that approaches (2) and (3) are typically by necessity two-step approaches, in which the bulk flow and the flow local to the RBCs are resolved separately. This is especially true for the BIM used here, as this method is only applicable to inertia-free flows. While this assumption is valid in the flow local to RBCs, it cannot be applied to large-sized flows where the Reynolds or Womersley numbers become large. Thus, the application of this method to complex geometries relies on a separate simulation to solve the Navier-Stokes equations in the macro-scale domain. } 

In our computational framework, the RBCs are discretized using spherical harmonic basis functions, which facilitate smooth representation of the RBC geometry and thus accurate evaluation of mechanical stresses within the RBC and the singular integrals that must be evaluated over the RBC surface. Some changes have been implemented in the solver, in particular the addition of cell-cell interactions and periodic boundary conditions. BIMs can utilize periodic boundary conditions efficiently, but some constraints must be placed on the problem to use such a method, which will be described later. This solver differs from previous works in multiple ways: for one, the complex fluid-structure interaction problem is solved directly, solving for both the motion of the fluid and membrane in conjunction. However, because of the BIM formulation, only the velocity on the surface of the cell is solved explicitly, reducing the dimensionality of the problem from three to two. Additionally, by necessity, the stress and strain history is tracked for each cell across all time steps. This information is of primary importance in hemolysis and calculating the likelihood that a cell will rupture.

The paper is outlined as follows: first, the methods for the solver are introduced, with particular attention paid to new elements of the solver, as well as validations of these new components for an ordered array of droplets. Next, results are presented and discussed for multiple cases, investigating the effects of the number of cells in the periodic box and the degree of the spherical harmonics representation, as well as comparisons between single-cell and multiple-cell simulations. Future work is discussed and finally, conclusions are presented.

% --------------------
\section{Materials and Methods}
The red blood cell motion is solved using a boundary integral solver introduced in a previous work \cite{rydquist_cell-resolved_2022}, the {finer} details of which are deferred to that paper. In brief, the RBCs are represented using a spherical harmonic expansion for each Cartesian component of the surface, which facilitates fast and accurate evaluation of the singular integrals in the boundary integral equation, a reformulation of the Stokes equations in integral form. The solid mechanics of the cell membrane are modeled using a combination of the Skalak model \cite{skalak_strain_1973} for the shear and dilation resistance and the Helfrich model \cite{helfrich_elastic_1973} for the bending resistance. 

With that earlier implementation, one may resolve the dynamics of a single cell in an unbounded flow that varies linearly at infinity based on a given velocity gradient tensor. To relax the isolated cell assumption and generalize that approach to model multiple cells, the current work requires two major additions to that legacy solver: interactions between cells and periodic boundary conditions, which are outlined in the following sections. Additionally, instead of directly solving the resultant system of equations, an iterative GMRES solver was used, as this resulted in a large reduction of computational cost and memory usage. 

In addition to the above change, the high dilatation ratio causes the equations to become stiff, particularly as the Capillary number is reduced to obtain lower shear rates. {The dilatation ratio is the ratio of the dilatation modulus, i.e., the RBC's resistance to local area changes, to the shear modulus.} While the overall cost of using a small enough time step is acceptable for single-cell simulations, the cost becomes prohibitive as the number of cells is increased. To combat the stiffness, a dilatation ratio of 20 is used. This is consistent with other works when accounting for the way this parameter is non-dimensionalized \cite{barthes-biesel_effect_2002, pozrikidis_numerical_2003, zhao_spectral_2010, sinha_dynamics_2015}, in which it has been shown to satisfactorily capture RBC dynamics.
% --------------------
\subsection{{Single-cell, Non-Periodic Formulation}}
{This section reviews the methodology for simulating a single cell in an unbounded flow presented in a previous work \cite{rydquist_cell-resolved_2022}. This earlier work also presents validations and results for the simulation of single RBCs. As mentioned above, the boundary integral method is used to simulate the motion of the cells. The boundary integral method uses a reformulation of the Stokes equations in integral form to solve for the velocity of the cell membrane and is derived using the reciprocal theorem to relate the flow of the RBC membrane to that of a point force. The cell membrane is represented as a two-dimensional surface with resistance to shear, bending, and area dilatation. For such a surface, the boundary integral equation is expressed as}

\begin{equation}\label{eq:oldBIM}
    {u_j(\mathbf x_0) - \frac{1-\lambda_{{v}}}{4\pi(1+\lambda_{{v}})}D_{ij}[u_i](\mathbf{x}_0-\mathbf x) = \frac{2}{1+\lambda_{{v}}}u^{\infty}_j(\mathbf x_0) - \frac{1}{4\pi\mu(1+\lambda_{{v}})}S_{ij}[\Delta f_i](\mathbf x_0-\mathbf x).}
\end{equation}
Here, $u_j$ is the velocity, $\mathbf x_0$ is the location where the equation is evaluated (which, for this equation, must be on the surface of the cell), $\lambda_{{v}}$ is the ratio of the viscosity of the fluid internal to the cells to the viscosity of the fluid external, $u_j^\infty$ is the velocity of the fluid in the absence of the cells, $\mu$ is the external fluid velocity, $\Delta f_i$ is the traction jump over the cell's membrane, and $S_{ij}$ and $D_{ij}$ are the single and double layer integral operators performed over the surface of the RBC $S$, defined as follows:
\begin{equation}
    \begin{split}
    S_{ij}[\psi](\mathbf{x}_0) &= \int_{S}G_{ij}(\mathbf x, \mathbf{x}_0) \psi(\mathbf x) \mathrm dS_x,\\
    {D_{ij}[\psi](\mathbf{x}_0)}&= \int_{S}T_{ijk}(\mathbf x, \mathbf{x}_0)\psi(\mathbf x) n_k(\mathbf x) \mathrm dS_x.
    \end{split}
\end{equation}
$G_{ij}$ is the free-space Green's function representing the velocity due to a point force, also known as the \textit{Stokeslet}, defined as
\begin{equation}
    G_{ij}(\mathbf x,\mathbf x_0) = \frac{\delta_{ij}}{r} +\frac{r_ir_j}{r^3},
\end{equation}
and $T_{ijk}$ is the free-space Green's function representing the stress field due to a point source, also known as the \textit{stresslet}, defined as
\begin{equation}
    T_{ijk}(\mathbf x,\mathbf x_0) = \-6\frac{r_ir_jr_k}{r^5},
\end{equation}
where $\mathbf r = \mathbf x - \mathbf x_0$ and $r=|\mathbf r|$. {Note that repeated indices in the above equations imply summation.} {Although not how the equation is derived, the boundary integral can be thought of as being somewhat similar to representing the RBC as a surface distribution of point forces. For a comprehensive review of the boundary integral method, readers are referred to the following text by Pozrikidis \cite{pozrikidis_boundary_1992}.}

{All quantities in Eq. \ref{eq:oldBIM} are known from material properties and information about the surrounding flow except for $\mathbf u(\mathbf x_0)$ (which is the unknown to be calculated from this equation) and $\Delta \mathbf f$. The latter quantity is obtained using the assumption that the membrane is inertia-less owing to the fact that it is thin and that the internal membrane load must be balanced by the traction jump. Thus, the internal membrane load is first calculated, which directly yields $\Delta \mathbf f$. The internal membrane load is obtained using a combination of the Skalak constitutive model \cite{skalak_strain_1973}, which is a strain-hardening model that represents the cell's resistance to shear and area dilatation, and the Helfrich bending model \cite{helfrich_elastic_1973}.}

{The Skalak model leads to the following equation for the surface tension tensor:}
\begin{equation}\label{eq:tau}
    {\boldsymbol{\tau} = G\left[\frac{1}{\lambda_1\lambda_2}\left(I_1 + 1\right)\mathbf V^2 + \lambda_1\lambda_2\left(CI_2 - 1\right)\left(\mathbf I - \mathbf{nn}\right)     \right].}
\end{equation}
{Here, $G$ is the membrane shear modulus, $C$ is the dilatation ratio, $\lambda_{1,2}$ are the local principal stretches, $\mathbf I$ is the identity tensor, $\mathbf n$ is the normal vector, $I_1 = \lambda_1^2 + \lambda_2^2 -2$ and $I_2=\lambda_1^2\lambda_2^2 - 1$ are surface strain invariants, and $\mathbf V^2$ is the left Cauchy-Green Deformation tensor. This equation can be calculated in local coordinates by comparing the current configuration of the RBC to the reference configuration, which in this case is an oblate spheroid, and the contribution to $\Delta \mathbf f$ from the tension can be obtained via an infinitesimal force balance as:}
\begin{equation}\label{eq:tens_tracs}
\begin{split}
    {\Delta f_t^\beta }&{= -\tau^{\alpha\beta}_{|\alpha},}\\
    {\Delta f_t^3} &{= -\tau^{\alpha\beta}b_{\alpha\beta}.}
\end{split}
\end{equation}
{$\alpha$ and $\beta$ here refer to the two local, tangent coordinate directions and $b_{\alpha\beta}$ is the curvature tensor. The contribution from bending ultimately yields}
\begin{equation}\label{eq:bend_trac}
    {\Delta\mathbf f_b = E_B\left[2\Delta_s\kappa + \left(2\kappa + c_0  \right)\left(2\kappa^2-\kappa_g-c_0 \kappa \right)\right]\mathbf n.}
\end{equation}
{$\Delta_s$ is the Laplace-Beltrami operator, $\kappa_g$ is the Gaussian curvature, $E_B$ is the bending modulus, $\kappa$ is the mean curvature, and $c_0$ is the spontaneous curvature.}

{As mentioned, the cell geometry is represented in terms of spherical harmonic basis functions. This parametrization takes the form:}
\begin{equation}\label{eq:xcoeffs}
    {\mathbf x(\theta,\phi) = \sum_{k=0}^m\sum_{l=-k}^k \hat{\mathbf x}_k^l Y_k^l(\theta,\phi).}
\end{equation}
{$\theta$ and $\phi$ are the inclination and azimuthal angles, respectively, in a spherical parameter space, and $Y_k^l(\theta,\phi)$ are spherical harmonic functions defined as}
\begin{equation}\label{eq:sph}
    {Y^l_{k}(\theta,\phi) = \sqrt{\frac{2k+1}{4\pi}\frac{(k-l)!}{(k+l)!}}P^l_{k}(\cos\theta)e^{il\phi}.}
\end{equation}
{$P_k^l$ are the associated Legendre polynomials.}

{The problem is non-dimensionalized primarily in terms of the external fluid viscosity $\mu$, the membrane shear modulus $G$, and the equivalent cell radius $a$, defined as the radius of a sphere with equivalent volume to the RBC. Other material properties are non-dimensionalized according to these parameters. The main parameter coupling the material properties of the cell to the flow is the Capillary number $Ca = \mu \dot\gamma a/G$, where $\dot \gamma$ is a characteristic shear rate of the flow.}

% --------------------
\subsection{Periodic Boundary Conditions}
The periodic form of the boundary integral equation for this problem is expressed as follows:
\begin{equation}\label{eq:fullbim}
    \begin{split}
    u_j(\mathbf x_0) - \frac{1-\lambda_{{v}}}{4\pi(1+\lambda_{{v}})}&\sum_{\gamma=1}^{N_c} \sum_{\mathbf p} D^\gamma_{ij}[u_i^p](\mathbf{x}_0-\mathbf x_p) = \\ &\frac{2}{1+\lambda_{{v}}}u^{\infty}_j(\mathbf x_0) - \frac{1}{4\pi\mu(1+\lambda_{{v}})}\sum_{\gamma=1}^{N_c} \sum_{\mathbf p}S^\gamma_{ij}[\Delta f_i^p](\mathbf x_0-\mathbf x_p).
    \end{split}
\end{equation}
Here, $N_c$ is the total number of cells in the primary periodic box, the sum over $\mathbf p$ is the sum over all periodic images of the primary periodic box {(for example, $\mathbf x_p$ refers to the coordinates of the cell membrane in the $p^{th}$ box), and $\gamma$ refers to the surface the integral is being taken over.}

Ideally, the sums over the surrounding periodic boxes in Eq. \ref{eq:fullbim} could be truncated to obtain an accurate approximation of the effects of the surrounding cells. However, due to the slow decay of the cell-cell interactions, these sums are only conditionally convergent. Instead of evaluating them directly to some truncation, Ewald summation techniques are used to efficiently and accurately evaluate the sums. The literature for Ewald sums is well developed \cite{pozrikidis_computation_1996, lindbo_spectrally_2010, marin_boundary_2012, af_klinteberg_fast_2014, af_klinteberg_fast_2017}, and will only briefly be described here. In short, the sums are split into a short-range sum that decays rapidly in real space, and a smooth, long-range sum that decays rapidly in reciprocal space. The \textit{Stokeslet} is split using the Hasimoto decomposition: 
\begin{equation}
    \sum_{\mathbf p} G_{ij}(\mathbf x {-\mathbf x_p}, \mathbf x_0) = \sum_{\mathbf p} G^R_{ij}(\mathbf x-\mathbf x_p, \mathbf x_0, \xi) + \sum_{\mathbf k\ne 0}G^F_{ij}(\mathbf x, \mathbf x_0, \mathbf k, \xi),
\end{equation}
where $G^R_{ij}(\mathbf x,\mathbf x_0, \xi)$ is the real, short-ranged component of the sum:
\begin{equation}
    G^R_{ij}(\mathbf x, \mathbf x_0, \xi) = \left(\mathrm{erfc}(\xi r) -\frac{2\xi r}{\sqrt{\pi}}e^{-\xi^2r^2} \right)\frac{\delta_{ij}}{r} + \left( \mathrm{erfc}(\xi r) + \frac{2\xi r}{\sqrt \pi}e^{-\xi^2r^2}\right)\frac{r_ir_j}{r^3},
\end{equation}
and $G^F_{ij}(\mathbf x, \mathbf x_0, \mathbf k, \xi)$ is the smooth, reciprocal-space component of the sum:
\begin{equation}\label{eq:Gf}
    G^F_{ij}(\mathbf x, \mathbf x_0, \mathbf k, \xi) = \frac{8\pi}{Vk^4}\left(1+\frac{k^2}{4\xi^2} \right)(k^2\delta_{ij} -k_ik_j)e^{-(k^2/4\xi^2 + i\mathbf k\cdot\mathbf r)}.
\end{equation}
Here, $\xi$ is a splitting parameter that determines the relative weights of the two sums with units of inverse length, $V$ is the volume of the primary periodic box, and the sum over $\mathbf k$ represents a sum over the reciprocal lattice. The basis vectors of this lattice $\mathbf b_i$ can be defined from the basis vectors of the real-space lattice $\mathbf a_i$ as $\mathbf b_i = \frac{2\pi}{V}\varepsilon_{ijk}\mathbf a_j\mathbf a_k$, where $\varepsilon_{ijk}$ is the permutation tensor. It should be noted that there will be no self-interaction term that will arise in the evaluation of these sums. The self-interaction term is added when the point where the sums are evaluated coincides with the location of one of the point forces in the sum. In such a case, the point force where the sums are evaluated needs to be removed from the sums in the primary box. This results in an additional term arising from the reciprocal space sum. While the discretization of the integrals in the present case essentially reduces to a sum of point forces, the sums are not actually evaluated at any of these point forces.

The \textit{stresslet} is defined similarly to the \textit{Stokeslet}. However, the decomposition introduced by Marin et al \cite{marin_boundary_2012} is used for its slightly faster convergence properties than other decompositions. It is defined as follows:
\begin{equation} \label{eq:Tsplit}
    \sum_{\mathbf p} T_{ijk}(\mathbf x {-\mathbf x_p}, \mathbf x_0)= \sum_{\mathbf p} T^R_{ijk}(\mathbf x-\mathbf x_p, \mathbf x_0, \xi) + \sum_{\mathbf k\ne 0}T^F_{ijk}(\mathbf x,\mathbf x_0,\mathbf k, \xi) -\frac{8\pi}{V}r_i\delta_{jk}.
\end{equation}
Note the inclusion of a linear component. This is included to ensure there is no mean flow through the periodic cell resulting from the \textit{stresslet} distribution \cite{zhao_spectral_2010, marin_boundary_2012, af_klinteberg_fast_2014}. The real-space component is
\begin{equation}
    \begin{split}
        T^R_{ijk}(\mathbf x-\mathbf x_p, \mathbf x_0, \xi) &= C(\xi, r)r_ir_jr_k + D(\xi,r)(\delta_{ij}r_k + \delta_{jk}r_i + \delta_{ki}r_j);\\
        C(\xi,r) &=-\frac{6}{r^5}\mathrm{erfc}(\xi r) - \left( \frac{12\xi}{r^4\sqrt \pi} + \frac{8\xi^3}{r^2\sqrt\pi} \right), \\
        D(\xi,r) &= \frac{4\xi^3}{\sqrt\pi}e^{-\xi^2r^2}, 
    \end{split}
\end{equation}
and the reciprocal space component is
\begin{equation}\label{eq:Tf}
    \begin{split}
        T^F_{ijk}(\mathbf x,\mathbf x_0,\mathbf k, \xi) &= -i\frac{8\pi}{V}A(\xi,\mathbf k)B(\mathbf k)e^{-(k^2/4\xi^2 + i\mathbf k\cdot\mathbf r)};\\
        A(\xi,\mathbf k) &= 1+\frac{k^2}{4\xi^2},\\
        B(\mathbf k) &= -\frac{2}{k^4}k_ik_jk_k + \frac{1}{k^2}(\delta_{ij}k_k + \delta_{jk}k_i + \delta_{ki}k_j).
    \end{split}
\end{equation}

While these sums are now convergent, they are still expensive to evaluate in their present form. However, the reciprocal-space sum can be accelerated by utilizing Fast Fourier Transforms (FFTs). The process involves first smearing the point forces to a uniform grid. FFTs are then used to transform this grid into reciprocal space, and a scaling representing Eqs. \ref{eq:Tf} and \ref{eq:Gf} is applied. This scaled version is then reverted back to real space and interpolated to the points where the sums must be evaluated. A more detailed description is given by Lindo and Tornberg \cite{af_klinteberg_fast_2017} for the Stokes equations and Deserno et al \cite{deserno_how_1998} for a more general description. Since the reciprocal-space sums can be evaluated faster than the real-space sums with this method, the value of $\xi$ is taken to be relatively large to shift the bulk of the work onto these sums.

Note that, due to the velocity term in the double layer operator, a linear velocity field must be used to ensure that the periodic images of a cell exactly replicate the cell in the primary box. This is because, to evaluate the periodic sums efficiently, the densities of the single and double layer operators ($\Delta f_i$ and $u_in_k$, respectively), must be equivalent between periodic images. At first, it may appear that even this is insufficient, as the velocities of the images are different from that of the primary cell, i.e., $u_i(\mathbf x_0) \ne u_i(\mathbf x_0 + \mathbf x_p)$. However, if the velocity field is linear, it can be assumed that the boxes are replicas of one another with the addition of a rigid body velocity, such that, for a constant velocity gradient $\nabla_i U_j$, $u_j(\mathbf x + \mathbf x_p) = u_j(\mathbf x) + \nabla_i U_j x_{p_i}$. Then, it is noted that the \textit{stresslet} has the property that 
\begin{equation} \label{eq:doub_props}
    \int_S T_{ijk}(\mathbf x, \mathbf x_0)n_k(\mathbf x)\mathrm d S_x = \delta_{ij}
    \begin{cases}
        4\pi, &\mathbf x_0 \in S,\\
        0, &\mathbf x_0 \text{ outside } S,\\
        8\pi, &\mathbf x_0 \text{ inside } S.
        
    \end{cases}
\end{equation}
Substituting the above velocity into the double layer integral for a cell in an arbitrary box $p$ evaluated at a point in the primary cell and assuming the shapes of the periodic images are equivalent to the shapes of the respective cell in the primary box yields
\begin{equation}
    \begin{split}
        \int_{S_p} T_{ijk}(\mathbf x - \mathbf x_p, \mathbf x_0) &u_i(\mathbf x - \mathbf x_p)n_k(\mathbf x - \mathbf x_p)\mathrm dS_x = \\
        =&\int_{S_p} T_{ijk}(\mathbf x - \mathbf x_p, \mathbf x_0) \left[u_i(\mathbf x) - \nabla_l U_i x_{p_l}\right]n_k(\mathbf x)\mathrm dS_x \\
        = &\int_{S_p} T_{ijk}(\mathbf x - \mathbf x_p, \mathbf x_0) u_i(\mathbf x)n_k(\mathbf x)\mathrm dS_x \\
        &- \nabla_l U_i x_{p_l}\int_{S_p} T_{ijk}(\mathbf x - \mathbf x_p, \mathbf x_0) n_k(\mathbf x)\mathrm dS_x\\
        =&\int_{S_p} T_{ijk}(\mathbf x - \mathbf x_p, \mathbf x_0) u_i(\mathbf x)n_k(\mathbf x)\mathrm dS_x,
    \end{split}
\end{equation}
thus ensuring that the double layer density is equivalent between boxes.

Additionally, minor changes were made to assist the solver with stability and accuracy. It was found that in some simulations the high relative translational velocities of the cells at different points in the box could introduce instability and inaccuracy. The velocity coefficients associated with this translational motion could be, in the worst cases, multiple orders of magnitude larger than the next largest velocity mode, and errors associated with the integration of this component could be large relative to these next highest modes. To mitigate this, the velocity on the surface of the cells was split into two components, $u_i(\mathbf x_0) = U_i + \tilde u_i(\mathbf x_0)$, where $U_i$ is the translational component of the velocity and $\tilde u_i(\mathbf x_0)$ is the portion of the velocity that varies over the cell surface. Substituting this into the double layer potential and using Eq. \ref{eq:doub_props} we find
\begin{equation} \label{eq:doub_sub}
    \begin{split}
        \int_S T_{ijk}(\mathbf x, \mathbf x_0) u_i(\mathbf x) n_k(\mathbf x)\mathrm dS_x &= U_i\int_S T_{ijk}(\mathbf x, \mathbf x_0)n_k(\mathbf x)\mathrm dS_x + \int_S T_{ijk}(\mathbf x, \mathbf x_0) \tilde u_i(\mathbf x) n_k(\mathbf x)\mathrm dS_x \\
        &= 
        \begin{cases}
            4\pi U_i + \int_S T_{ijk}(\mathbf x, \mathbf x_0) \tilde u_i(\mathbf x) n_k(\mathbf x)\mathrm dS_x, &\mathbf x_0 \in S,\\
            \int_S T_{ijk}(\mathbf x, \mathbf x_0) \tilde u_i(\mathbf x) n_k(\mathbf x)\mathrm dS_x, &\mathbf x_0 \notin S.
        \end{cases}
    \end{split}
\end{equation}
Substituting Eq. \ref{eq:doub_sub} into the left hand side of Eq. \ref{eq:fullbim} yields, for a surface $S$ containing the point $\mathbf x_0$,
\begin{equation}\label{eq:new_LHS}
    \begin{split}
    u_j(\mathbf x_0) - \frac{1-\lambda_{{v}}}{4\pi(1+\lambda_{{v}})}&\sum_{\gamma=1}^{N_c} \sum_{\mathbf p} D^\gamma_{ij}[u_i](\mathbf{x}_0-\mathbf x_p) = \\&\frac{2}{1+\lambda_{{v}}} U_j^S + \tilde u_j(\mathbf x_0) - \frac{1-\lambda_{{v}}}{4\pi(1+\lambda_{{v}})}\sum_{\gamma=1}^{N_c} \sum_{\mathbf p} D^\gamma_{ij}[\tilde u_i](\mathbf{x}_0-\mathbf x_p),
    \end{split}
\end{equation}
where $U_j^S$ is the translational velocity of surface $S$. Thus, integrals are only calculated over the spatially varying component of the velocity. This is easily facilitated by the spherical harmonic representation of the velocity, as the translational component of the velocity is represented solely by the $n,m=0$ modes of the spherical harmonic expansion.

% --------------------
\subsection{Singular and Near-singular Integration}
In the previous work, properties of spherical harmonics were used to construct a spectrally accurate integration rule based on rotating the spherical harmonic coefficients such that the pole of the coordinate system aligned with the singularity. However, this method is no longer applicable when the evaluation point is near, but not on, the surface, and it also is not applicable for use with the split Green's functions introduced to evaluate the periodic sums. As a result, a different method is required to calculate all singular and near-singular integrals. The partition of unity method used by Zhao et al \cite{zhao_spectral_2010} was used. This method splits the singular integrals into two components:

\begin{equation} \label{eq:sing_rule}
    \int_S G_{ij}(\mathbf x, \mathbf x_0) \phi(\mathbf x)\mathrm dS_x = \int_SG_{ij}(\mathbf x, \mathbf x_0)\eta(f) \phi(\mathbf x)\mathrm dS_x + \int_S G_{ij}(\mathbf x, \mathbf x_0)[1-\eta(f)] \phi(\mathbf x)\mathrm dS_x
\end{equation}
where
\begin{equation}
    \eta(f) =
    \begin{cases}
        \exp{\left(\frac{2e^{-1/f}}{f-1}\right)}, & f=\rho/\rho_1<1,\\
        0, &{\rm otherwise}.
    \end{cases}
\end{equation}
$\rho$ is the great circle distance in the reference unit sphere parameter space from $(\theta,\phi)$ to $(\theta_0,\phi_0)$, where these are the respective coordinates used to parametrize $\mathbf x$ and $\mathbf x_0$. Here, $\rho_1$ is a cutoff parameter defined as $\rho_1 = \pi/\sqrt{{m}}$, outside of which the first integral on the right-hand side of Eq. \ref{eq:sing_rule} does not have support. ${m}$ is the degree of the spherical harmonic expansion used to represent the cell shape.

The second component is smooth over the surface of the cell, and can thus be evaluated using a normal, non-singular integration rule. The first component is rapidly varying near the location of the evaluation point $\mathbf x_0$ but only has support over a small patch in parameter space. To accurately evaluate this integral, a $\sinh$ transformation is used to group points near $\mathbf x_0$ \cite{johnston_sinh_2005}. In the case of a near-singular integrand, the point on the upsampled grid used for anti-aliasing that is closest to $\mathbf x_0$ on the integration surface is used for the center of the patch.

% --------------------
\subsection{Close-range Interactions}
RBCs have several adhesive and repulsive interactions with each other at close range. These interactions are often modeled all at once via a Morse potential \cite{liu_coupling_2004, wang_numerical_2009, li_computational_2014, xiao_simulation_2016}. However, these interactions are relatively weak compared to the fluid forces at the shear rates anticipated for use with the solver. For example, these attracting forces are thought to be responsible for the rheological properties of a suspension of RBCs at low` shear rate, while at high shear rate, the prime driver of the rheological properties is the RBC deformation {\cite{chien_blood_1967, chien_shear_1970}}. Therefore, these interactions are not included. However, a short-range Lennard-Jones force has been included in some simulations \cite{javadi_silico_2021}. {This force is strongly repulsive at close range, and is utilized to prevent the cells from overlapping, instead of representing a physical force. This force is selected because it becomes very strongly repulsive at short range, but also decays very quickly as the cells separate}. Hydrodynamic interactions typically prevent this overlap, but it can still occur on occasion, so this force is included in the current work. 

{Note that the inclusion of this repulsive force is a numerical treatment and not meant to mimic any physical process. As such, the resultant force is not incorporated into the calculation of the effective viscosity below. Other methods of ensuring the RBC surfaces do not cross, for example, include simply moving the membranes a fixed distance from each other if they come into too close of contact \cite{zhao_spectral_2010}.}

The interaction energy associated with the Lennard-Jones force is given by
\begin{equation}
    U_{LJ}(r) = 4\varepsilon\left[ \left(\frac{\sigma}{r} \right)^{12} - \left(\frac{\sigma}{r} \right)^{6}\right].
\end{equation}
In the above equations, $\varepsilon$ is a scaling constant for the energy and $\sigma$ is a scaling constant for the separation distance $r$. The force is then calculated as $f = \partial U /\partial r$, which is supplemented to the traction jump. {This force is only included when the integration surface is not on the same surface as the evaluation point to avoid interactions between adjacent grid points. When performing integrals over the surface of the cell, this force is supplemented to the traction jump if the distance between a given integration point and the evaluation point is less than the cutoff distance.}

This repulsive force is also useful in setting a minimum distance between the cells, as the accuracy of the integration rule degrades as the cells come into close contact. This interaction energy vanishes at $2^{1/6}\sigma$ and rapidly increases as the cells draw nearer. The integrals become increasingly difficult as the spacing between the cells becomes smaller than the average cell spacing, defined as $h = \sqrt{A}/({m}+1)$, with particularly large error around $h/2$. As such a value of approximately $\sigma = h/2^{7/6}$ is used. However, test simulations have shown minimal differences in simulations with and without this force when the cells do not overlap. {When comparing a case where this force is included to one where it is not and the cells do not overlap, a difference in calculated effective viscosity of 1.23\% has been observed. In addition, the average difference of the parameter $\max(\lambda_1/\lambda_2)$, a measure of maximum shear defined in greater detail below, is 2.26\% between the two cases. These simulations were performed at a hematocrit of 25\% and a shear rate of 1000 $s^{-1}$, and while these differences are not particularly large at this hematocrit, it is possible that the force could be responsible for larger differences at greater hematocrits, as the cells are forced into closer contact.}

% --------------------
\subsection{{Numerical Implementation}}
{The numerical procedure for solving the problem is described as follows:}
\begin{enumerate}
    \item {The spherical harmonic and geometric quantities are pre-computed.}
    \item {The time step loop is started.}
    \item {The value of $\Delta \mathbf f$ is calculated for all cells.}
    \item {The GMRES loop is started, using the velocity of the previous time step as an initial guess for the velocity.}
    \item {For a given GMRES iteration, the real-space components of Eq. \ref{eq:fullbim} are calculated for all cells and evaluation points.}
    \item {The reciprocal-space components of Eq. \ref{eq:fullbim} are calculated and added to the real-space components.}
    \item {The residual is calculated, and iterations are run until a specified tolerance is reached.}
    \item {The cell membrane and periodic basis vectors are advanced. If necessary, the basis vectors are reparametrized.}
    \item {The time step loop is iterated.}
\end{enumerate}

% --------------------
\section{Results} \label{sec:results}
Several facets of the solver were investigated. Ultimately, the goal of the current work is to investigate the cheapest method of obtaining a faithful reconstruction of the cells' behavior. To this end, simulations were run investigating differences not just between periodic simulations and non-periodic simulations, but also between periodic simulations at varying periodic box length and mesh resolution.

% --------------------
\subsection{Validation}
\begin{figure}[htbp]
  \centering
  \subfloat[ \label{sub:drops}]{{\includegraphics[height=0.3\textwidth]{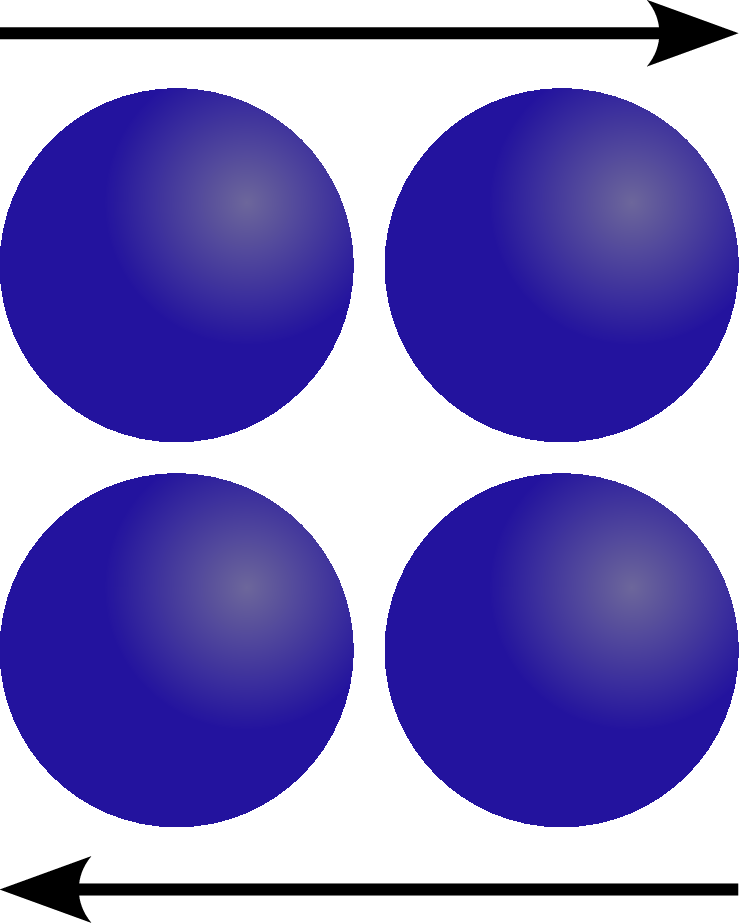}}}\qquad%
  \subfloat[ \label{sub:zinch}]{{\includegraphics[scale=0.42]{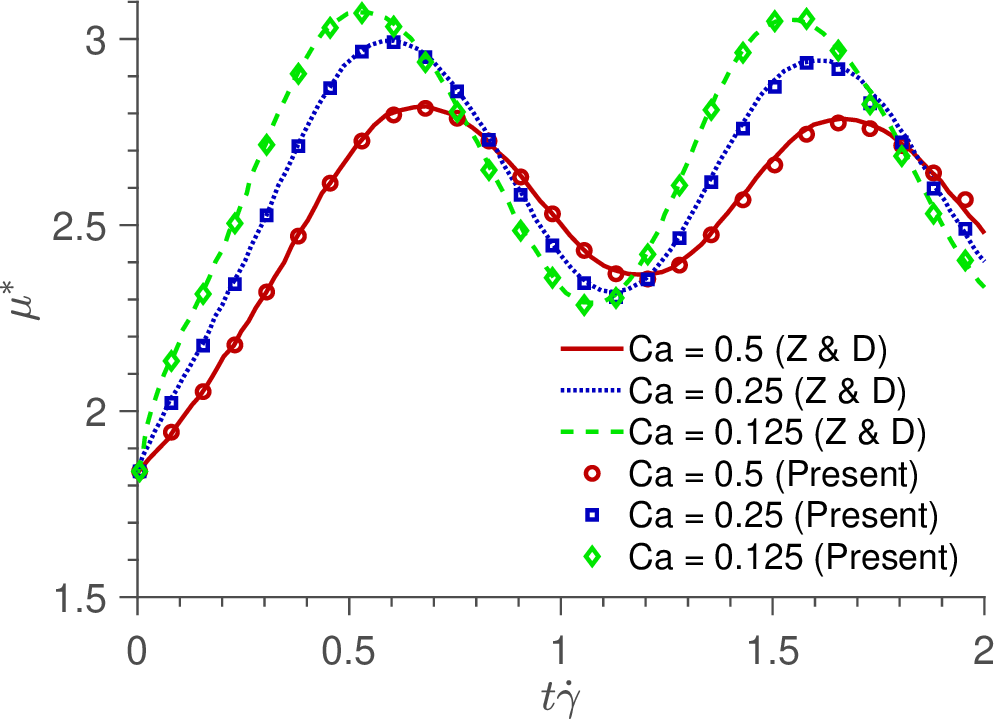}}}
  \caption{Results for a periodic array of droplets in shear flow (a). (b) The calculated effective viscosity of the suspension over time, for three different values of the Capillary number, effectively a measurement of the strength of the shear flow. Results of our computations (markers) are compared with those produced by Zinchenko and Davis \cite{zinchenko_shear_2002} (lines), showing good agreement.}
  \label{fig:drops}
\end{figure}

{The bulk of the boundary integral method solver is validated in our earlier article \cite{rydquist_cell-resolved_2022}. To ensure the validity of the new additions to the solver, in particular the periodic components, we validate it here against simulations by Zinchenko and Davis \cite{zinchenko_shear_2002}.} These simulations are performed for an ordered suspension of droplets in shear flow that only have {surface tension}. The relative viscosity of the suspension to the ambient fluid is shown in Fig. \ref{fig:drops} for three values of the Capillary number, which, in this case, measures the relative strength of {the surface tension} to the strength of the shear flow. Note that for shear flow, the effective viscosity in a fully non-dimensional form is computed as \cite{zinchenko_shear_2002, batchelor_stress_1970}
\begin{equation}\label{eq:eff_visc}
    \mu^* = 1 + \frac{1}{V}\sum_{\gamma=1}^{N_c}\int_{S_\gamma}{\left[\Delta f_1 x_2 + (\lambda_{{v}} - 1)(u_1n_2 + u_2n_1) \right]}\mathrm dS_\gamma.
\end{equation}
Here, $\mu^*$ is the viscosity of the suspension relative to the viscosity of the surrounding fluid, $V$ is the volume of the periodic box and $n_i$ is the normal vector. {In practice, this equation is derived by calculating the shearing component of the average stress tensor aligned with the shear flow divided by the viscosity of the surrounding fluid, which is why only the $1,2$ components are utilized.} In this case, 8 droplets in the primary cell were simulated, with a spherical harmonic truncation of ${m}=12$ used. This case was simulated with a volume fraction of $0.95(\pi/6)$ and a viscosity ratio of $\lambda_{{v}} =5$. Excellent agreement is obtained between the present solver and that of Zinchenko and Davis at three different Capillary numbers of 0.125, 0.25, and 0.5.

% --------------------
\subsection{Shear flow simulations setup}
\begin{figure}[H]
    \centering
    \subfloat[ \label{sub:RBCs}] {{\includegraphics[width=0.42\textwidth]{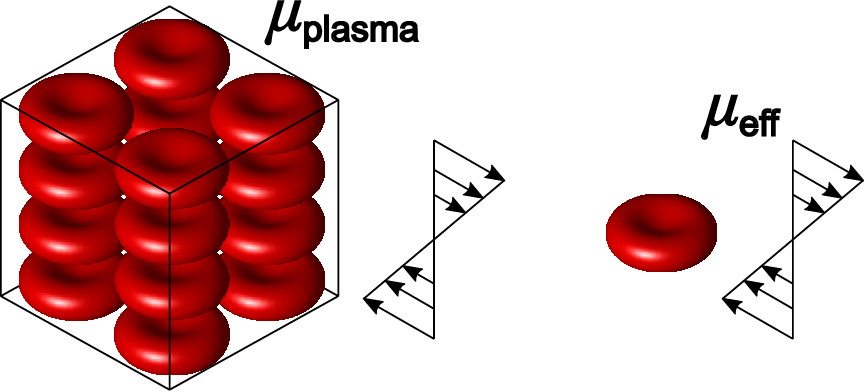}}}\qquad%
    \subfloat[ \label{sub:colored500}] {\includegraphics[scale=0.42]{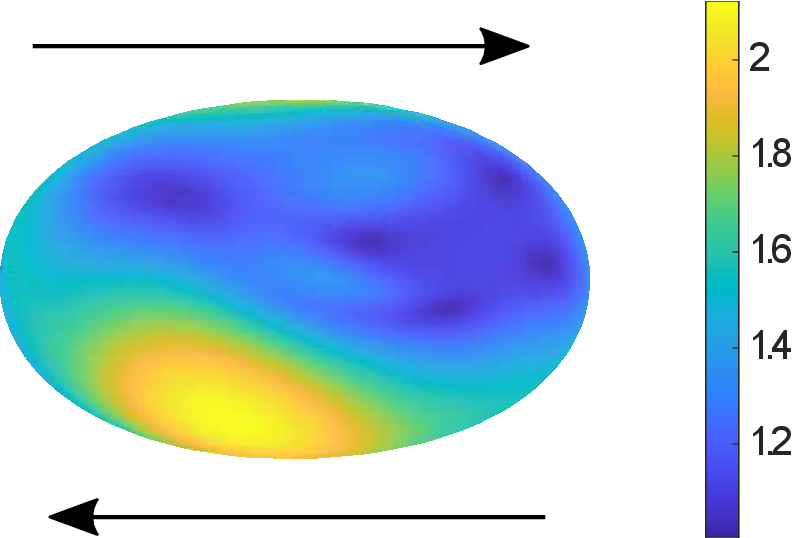}}\\
    \subfloat[ \label{sub:spindle}] {\includegraphics[scale=0.6]{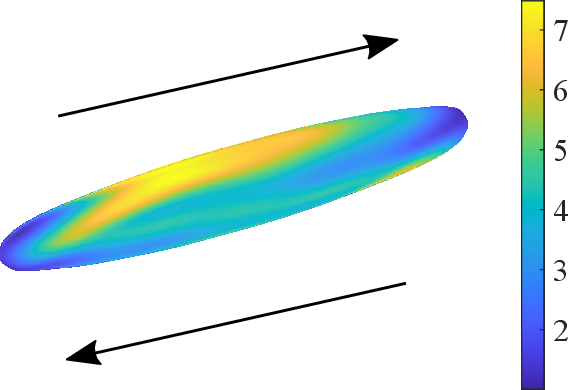}}\qquad 
    \subfloat[ \label{sub:trayson}] {\includegraphics[scale=0.6]{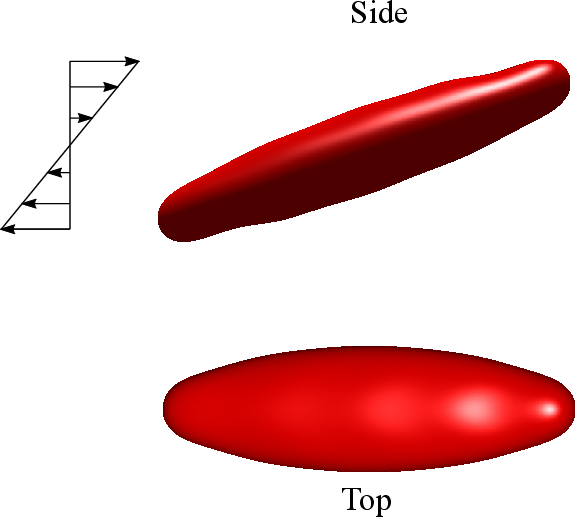}} 
    \caption{(a) The two main cases under consideration: multiple interacting cells in a box with periodic boundary conditions (left) and single-cell simulations using an effective viscosity corresponding to the shear rate and hematocrit of the corresponding multiple-cell case (right). (b) A contour plot of the value $\lambda_1/\lambda_2$ on the surface of a RBC in shear flow, representing the shear strain the cell is undergoing at a specific point in time. Of particular interest is the maximum value of this parameter, where it is presumed hemolytic events are most likely to occur. {(c) A RBC in a high-viscosity, high-shear flow, roughly correlating to results from Sohrabi \& Liu \cite{sohrabi_cellular_2017}. (d) A RBC in a shear flow of shear rate $\dot\gamma = 200$ $s^{-1}$ and surrounding fluid viscosity 30 cP. This simulation roughly correlates to results from Tran-Son-Tay et al \cite{tran-son-tay_membrane_1987}.}}
    \label{fig:setup}
\end{figure}

Given that the RBC damage often occurs in high shear flow environment, we consider an unbounded shear flow to establish the difference between single- and multiple-cell simulation approaches. {Shear flow is also selected due to the relative simplicity in calculating the effective viscosity, whereas establishing this viscosity in other flows is not as straightforward. Additionally, implementing periodic boundary conditions for this flow type is fairly trivial. Extensional flow has been shown to induce large deformations in cells \cite{faghih_deformation_2020}; however, these types of flows were not investigated in this work. Periodic boundary conditions can be implemented in extensional flows in a fairly straightforward manner, but initial simulations were largely unstable over long time periods. Further work is required to extend this work to extensional or arbitrary flow types.} Figure \ref{sub:RBCs} shows a schematic of the {shear flow} approaches, with the multiple-cell, periodic boundary condition approach on the left, and the single-cell, effective viscosity approach on the right. Before doing so, however, we first establish the minimum mesh resolution and periodic box size for accurate multiple-cell simulations in the following two subsections. 

The main parameter of interest investigated in the following simulations is $\lambda_1/\lambda_2$, where $\lambda_{1,2}$ are the principal {stretches} at a given point in time on the surface of the cell. {These parameters represent the ratio of the dimensions of the local area element in the deformed state to the reference shape, and $\lambda_1/\lambda_2$ is an invariant of these ratios that represents the shear deformation. A} value of 1 represents the membrane in its unstrained configuration. We report the maximum value of $\lambda_1/\lambda_2$ over the entire cell, as it is presumed that hemolytic events would occur at these locations. {This parameter has been used in previous works \cite{lu_boundary_2019, zhu_dynamics_2022}, as it provides one measure of the maximum deformation the RBC membrane is undergoing. Additionally, the strain is useful because hemolysis in cell-resolved simulations is typically examined in terms of local strain \cite{xu_cell-scale_2023, nikfar_multiscale_2020, sohrabi_cellular_2017}.} The average value of this parameter is also examined as a second measure of the similarity between different types of simulation. Additionally, since these values vary over time, their time averages are used here to compare across different simulation cases. A contour plot of $\lambda_1/\lambda_2$ of a cell in shear flow is shown in Fig. \ref{sub:colored500}. In this figure, the maximum value occurs on the sides of the cell, as this location is stretched most when the cell aligns with the shear flow.
{Figures \ref{sub:spindle} and \ref{sub:trayson} display some of the behavior of RBCs under extreme conditions. These simulations do not correspond to simulations performed in this study as they do not correspond to physiological conditions or are well above the regime where blood is shear-thinning, but they are valuable as they correspond to some scenarios available in the literature. These simulations are performed using a relatively higher value of the surrounding fluid viscosity and shear rate than the scenarios under investigation in this study. Under these circumstances, the cells take on an elliptical steady-state shape \cite{sohrabi_cellular_2017, sutera_deformation_1975, faghih_deformation_2020, tran-son-tay_membrane_1987}. Qualitatively the shapes of the cells in these figures match well with those shown in the referenced studies. The map of the strain in Fig. \ref{sub:spindle} matches well with that produced by Sohrabi \& Liu \cite{sohrabi_cellular_2017}, and the general shape shown in Fig. \ref{sub:trayson} matches well qualitatively with those imaged by Tray-Son-Tay et al \cite{tran-son-tay_membrane_1987}.}

\begin{figure}[H]
   \centering
   \includegraphics[scale=.42]{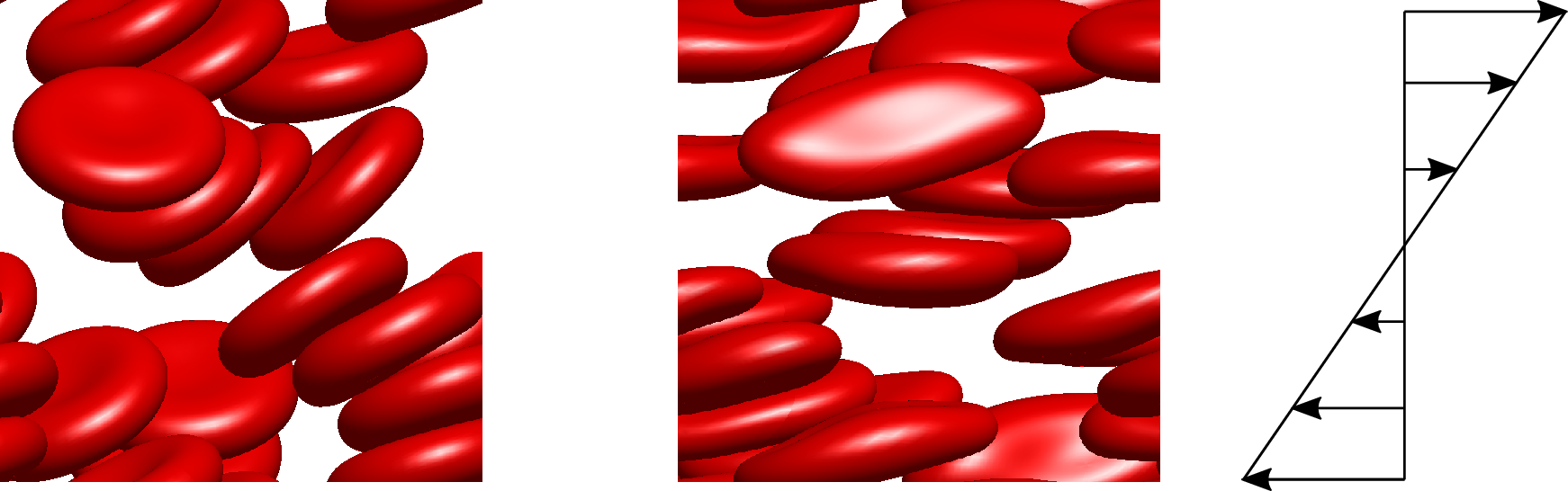}
   \caption{{The initial configuration of the RBCs prior to exposing them to a constant shear rate (left) and the cells after being exposed to a shear rate of 1000 $s^{-1}$ (right).}}
   \label{fig:RBC_side}
\end{figure}

{Initially, simulations were performed using ordered setups similar to the periodic simulation shown in Fig. \ref{sub:RBCs}. However, it was found that this configuration could lead to minor changes in the calculated effective viscosity and $\max(\lambda_1/\lambda_2)$ from cells that were initially randomly distributed, even after an instability caused the cells to deviate from their initial ordered configuration. As such, before the shear flow is started, the cells are individually subjected to random velocities for several seconds so they are able to reach a more random distribution in the periodic box. A side view of this simulation at the start is displayed in Fig. \ref{fig:RBC_side}, as well as a side view of the cells after being exposed to a shear rate of 1000 $s^{-1}$.}
% end at multi "80.3" 1K_16

\begin{figure}[H]
   \centering
   \includegraphics[scale=.2]{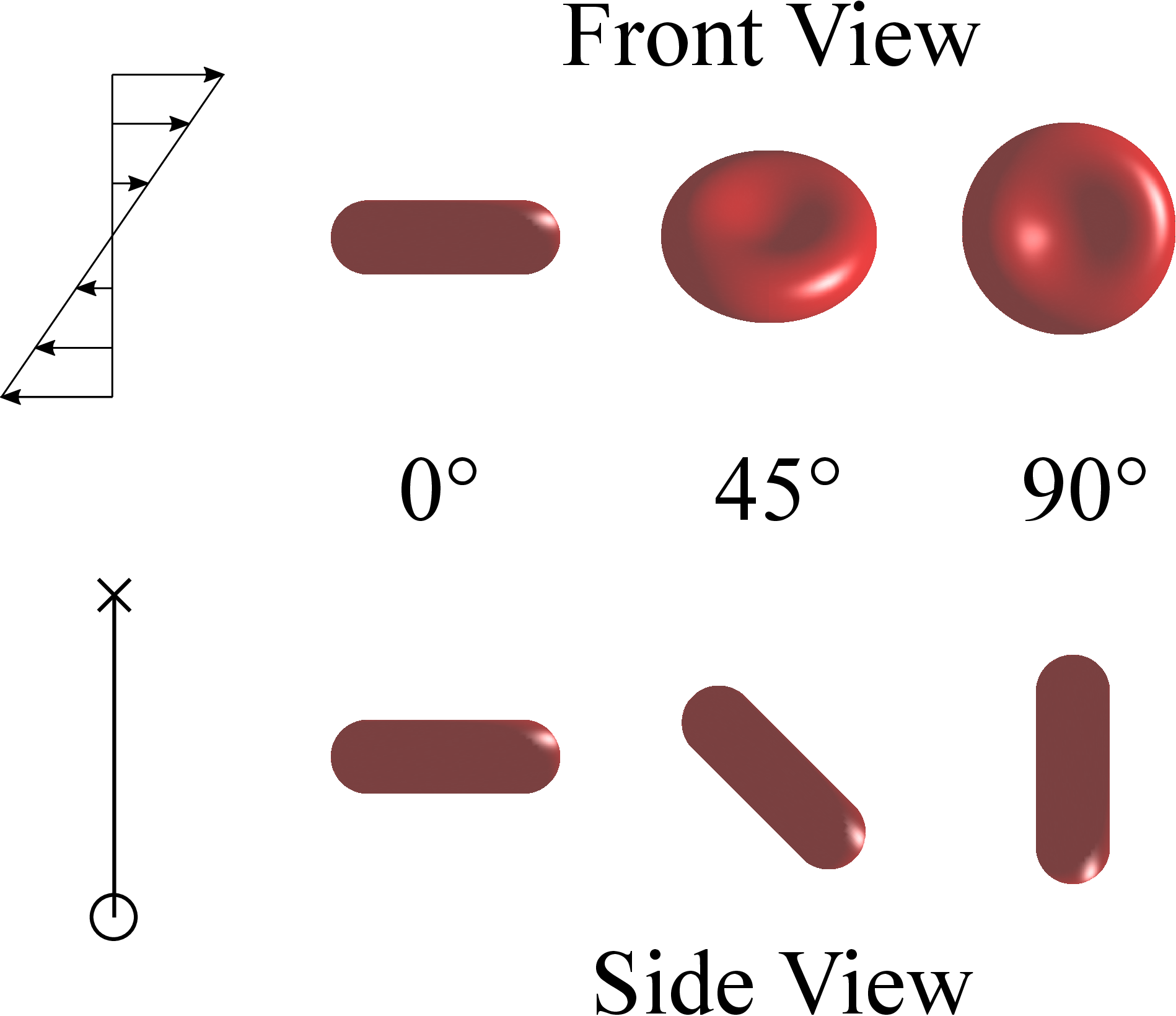}
   \caption{{The three initial orientations the single-cell simulations are run from. In the below results, the average of these three orientations is typically compared to the multiple-cell simulations.}}
   \label{fig:Orientations}
\end{figure}

{Also, the RBCs in the single-cell simulations can take on different steady-state behaviors depending on the initial configuration of the cell relative to the shear flow. As a result, the single-cell simulations are performed for three initial orientations, and most of the results in the below sections are compared against the average of these three orientations. Figure \ref{fig:Orientations} displays these three orientations, which are named in terms of the angle of the cell relative to the shearing direction.}

% --------------------
\subsection{Convergence studies}
\subsubsection{Mesh convergence}
\begin{figure}[H]
   \centering
   \includegraphics[scale=.42]{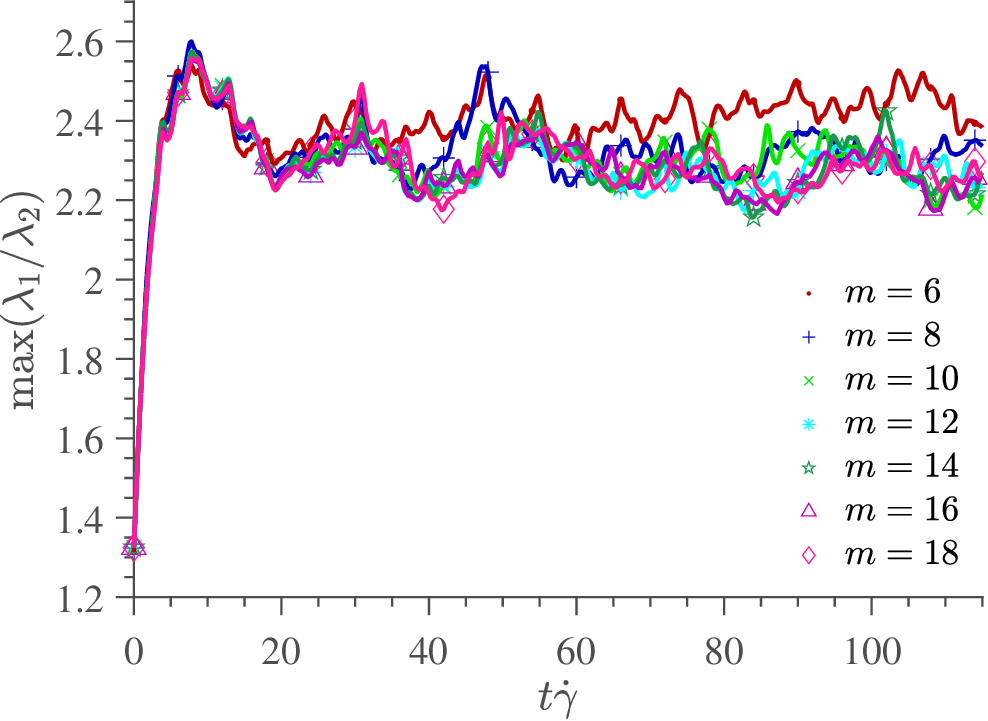}
   \caption{{The time history of $\max(\lambda_1/\lambda_2)$, averaged across all cells, plotted for all values of $m$. The $m=6$ and $8$ are the largest outliers in this figure.}}
   \label{fig:pplots}
\end{figure}

First, simulations were run to establish an appropriate truncation of the spherical harmonic expansion for the RBC geometry for the multiple-cell simulations. For this case, as well as cases in the following sections, a hematocrit of 25\% was utilized. Additionally, a shear rate of $\dot \gamma = 1000$ $s^{-1}$ was used. Simulations were performed using 16 cells. {The time history of $\max(\lambda_1/\lambda_2)$ for these cases, averaged across all 16 cells, is displayed in Fig. \ref{fig:pplots}} Results, displayed in Table \ref{tab:l1l2_p}, are presented for the time-average of the maximum value of $\lambda_1/\lambda_2$ over the surface of the cell, expressed in terms of the percentage difference from the ${m}=18$ case.  For the ${m}=18$ case, this value was $|| \max(\lambda_1/\lambda_2)|| = 2.292$. {Additionally, results for the effective viscosity as a function of maximum degree are displayed as well, with a relative viscosity at $m=18$ of $\mu^* = 1.482$.}

\begin{table}[ht!]
 \caption{The time-average of the maximum value of $\lambda_1/\lambda_2$ over the surface of the cell {and effective viscosity $\mu^*$}  as a function of the maximum degree used for the spherical harmonic basis function ${m}$ for the $\gamma = 1000$ ${s}^{-1}$ case. Results are expressed as a percentage difference from the ${m}=18$ case.}
\label{tab:l1l2_p}
 \centering
\begin{tabular}{ |p{3.3cm}||p{1.2cm}|p{1.2cm}|p{1.4cm}|p{1.4cm}|p{1.2cm}|p{1.2cm}|  }
 \hline
 ${m}$& 6 &8 &10 &12 &14 &16\\
 \hline
 {\% diff $\max(\lambda_1/\lambda_2)$}   &5.18    &1.17 &0.0331 &0.569 &0.402 &0.973  \\
 {\% diff $\mu^*$}   &6.47    &2.01 &0.00815 &0.00113 &0.0916 &0.310   \\
 \hline
\end{tabular}
\end{table}

Note that exact convergence is not reached, because minor differences in the simulation setup, including the truncation, can lead to minor differences in behavior. However, less than a 1\% difference between cases is reached at ${m}=10$ and above, so this truncation is used for the following cases. This convergence analysis was performed in the previous work for the single-cell simulations, showing that ${m}=10$ is sufficiently large for the convergence of single-cell simulations \cite{rydquist_cell-resolved_2022}.  {However, as with the simulations in the previous study, a value of $m=12$ is used for all simulations to ensure convergence is reached.}

% --------------------
\subsubsection{Effect of box size}
\begin{figure}[H]
    \centering
    \subfloat[ \label{sub:l1l2_pU}] {{\includegraphics[scale=0.42]{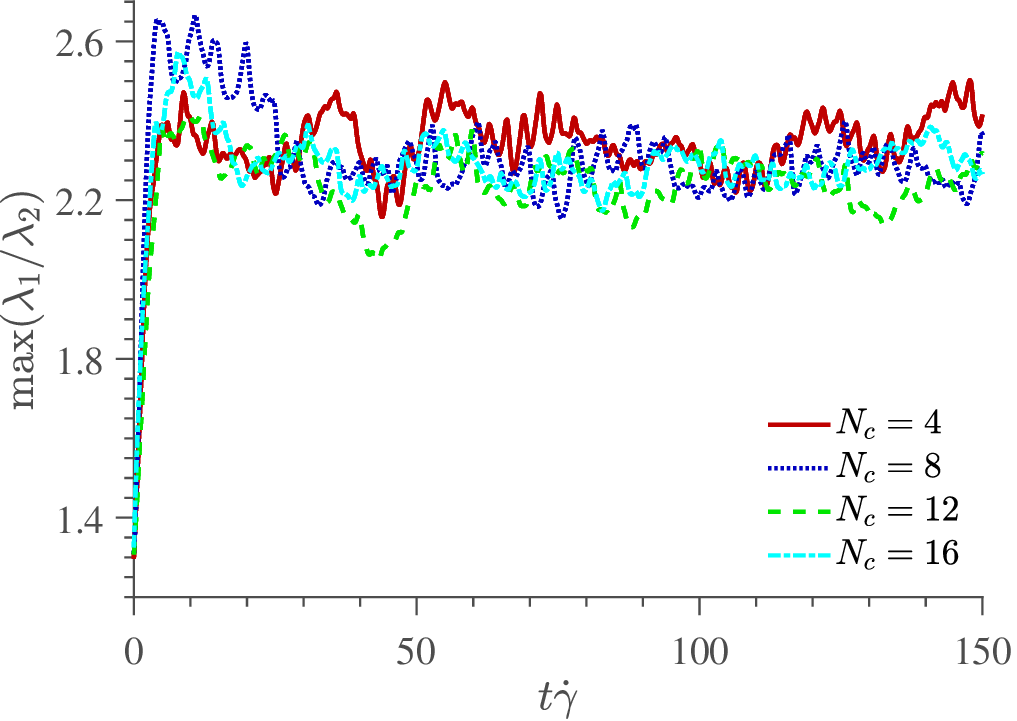}}}\qquad%
    \subfloat[ \label{sub:spectra}] {\includegraphics[scale=0.42]{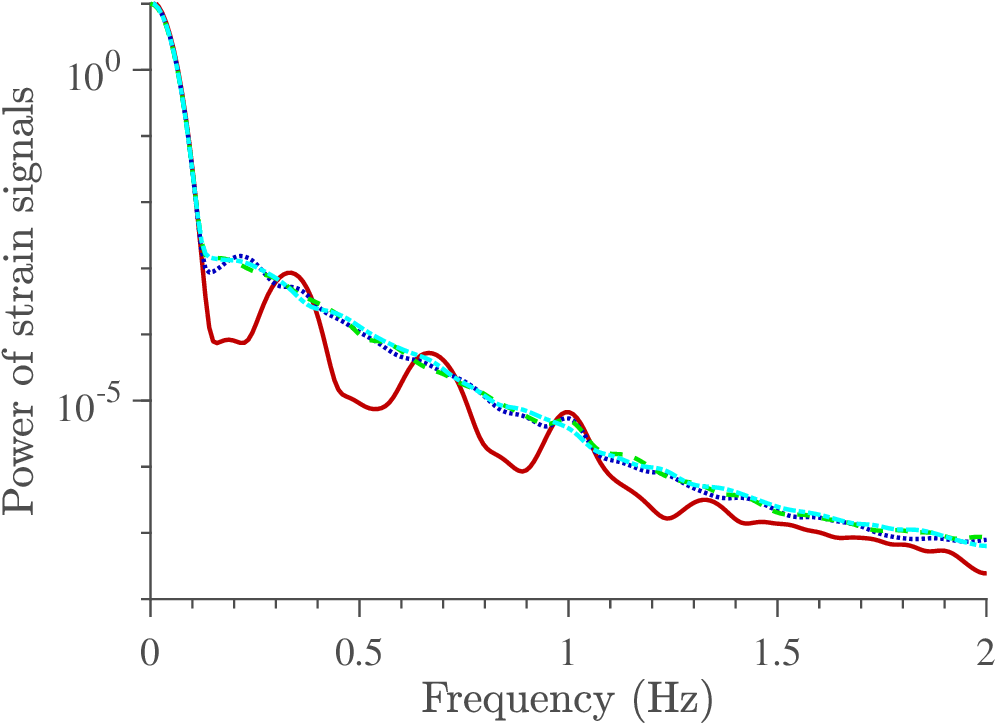}}
    \caption{Results of the simulations where the box size is varied. (a) The maximum value of the parameter $\lambda_1/\lambda_2$ on the surface of the cell, representing the shear strain, as a function of non-dimensional time for RBCs in a shear flow of strength 1000 $s^{-1}$. Results are shown for cases run with 4, 8, 12, and 16 cells. (b) The average of the spectral density of the individual signals. The signals show similar behavior, with notable differences for the $N_c={4}$ case, where the frequency is concentrated at specific points. However, the magnitude of these frequencies match well with the other cases.}
    \label{fig:boxconv}
\end{figure}

To investigate the effect of periodic box size on the results, we considered four cases containing 4 to 16 cells. For each case, we adjust the box size so that the hematocrit level remains fixed at 25\%. Simulating a single cell in periodic boundary conditions was tested as well, but these simulations were generally unstable as the cell would have close-range interactions with itself. These computations are performed at a shear rate of $\dot \gamma =1000$ ${s}^{-1}$. The maximum value of the parameter $\lambda_1/\lambda_2$ as a function of non-dimensional time is displayed in Fig. \ref{sub:l1l2_pU} for the cases of 4, 8, 12, and 16 cells superimposed on each other. {Results are presented in Table \ref{tab:l1l2_Nc} for percentages difference from the $N_c=16$ case.} Across the range of cells tested, the time average of the maximum value of this parameter varied by at most approximately {2.7\% when considering all cases and taking the $p=16$ case as the reference}. To further investigate the differences between these cases, however, the frequency content of these cases was examined as well. Figure \ref{sub:spectra} displays the spectral density of the maximum value of the $\lambda_1/\lambda_2$ signals in Fig. \ref{sub:l1l2_pU}.
These plots are similar for the most part, but show some differences, particularly between the $N_c={4}$ case and other cases. Specifically, the other cases appear to contain a broader and more uniform spectrum, whereas the $N_c={4}$ case appears to have a spectrum concentrated at specific frequencies. {These differences are perhaps a result of there not being enough cells to obtain a well-mixed solution, with the results being heavily influenced by the individual cells. Although the $N_c=8$, $12$, and $16$ cases appear to match both qualitatively and quantitatively fairly well, out of caution, the simulations in the following sections are run using 16 cells. Note that the effective viscosities of each of these cases were calculated as well, showing a maximum difference of approximately 2.3\% from the $N_c = 16$ case at $N_c=8$.}

\begin{table}[ht!]
\caption{The time-average of the maximum value of $\lambda_1/\lambda_2$ over the surface of the cell and effective viscosity $\mu^*$ as a function of the number of cells in the periodic box ${N_c}$ for the $\gamma = 1000$ ${s}^{-1}$ case. Results are expressed as a percentage difference from the $N_c=16$ case.}
\label{tab:l1l2_Nc}
 \centering
\begin{tabular}{ |p{3.3cm}||p{1.5cm}|p{1.5cm}|p{1.5cm}|  }
 \hline
 $N_c$& 4 &8 &12 \\
 \hline
 {\% diff $\max(\lambda_1/\lambda_2)$}   &2.66    &0.332 &1.78  \\
 {\% diff $\mu^*$}   &0.648    &2.25 &1.07 \\
 \hline
\end{tabular}
\end{table}

% --------------------
\subsection{Comparison of single- and multiple-cell simulations}
The underpinning assumption of the single-cell modeling approach is that the effect of cell-cell interactions on RBC dynamics can be captured by using the blood bulk viscosity rather than that of plasma. Thus, we first establish the bulk viscosity of the RBC suspension before comparing single- and multiple-cell simulations in the following subsections. 

\subsubsection{Computation of the bulk viscosity}
\begin{figure}[H]
    \centering
    \subfloat[ \label{sub:mu_eff1}] {{\includegraphics[scale=0.42]{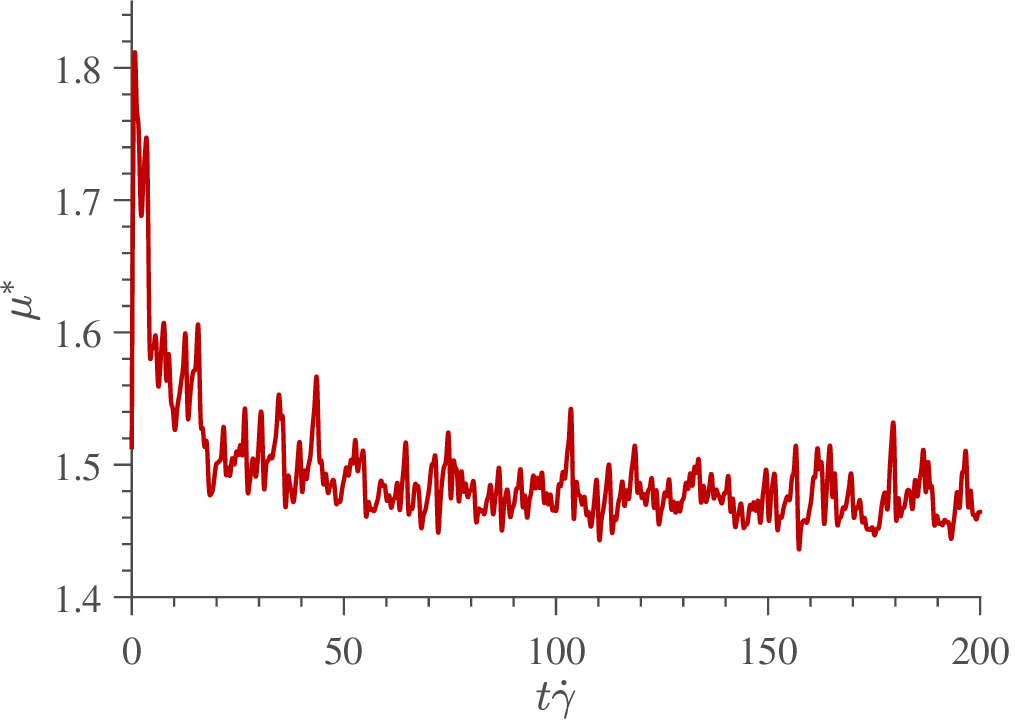}}}\qquad%
    \subfloat[ \label{sub:eff_visc}] {\includegraphics[scale=0.42]{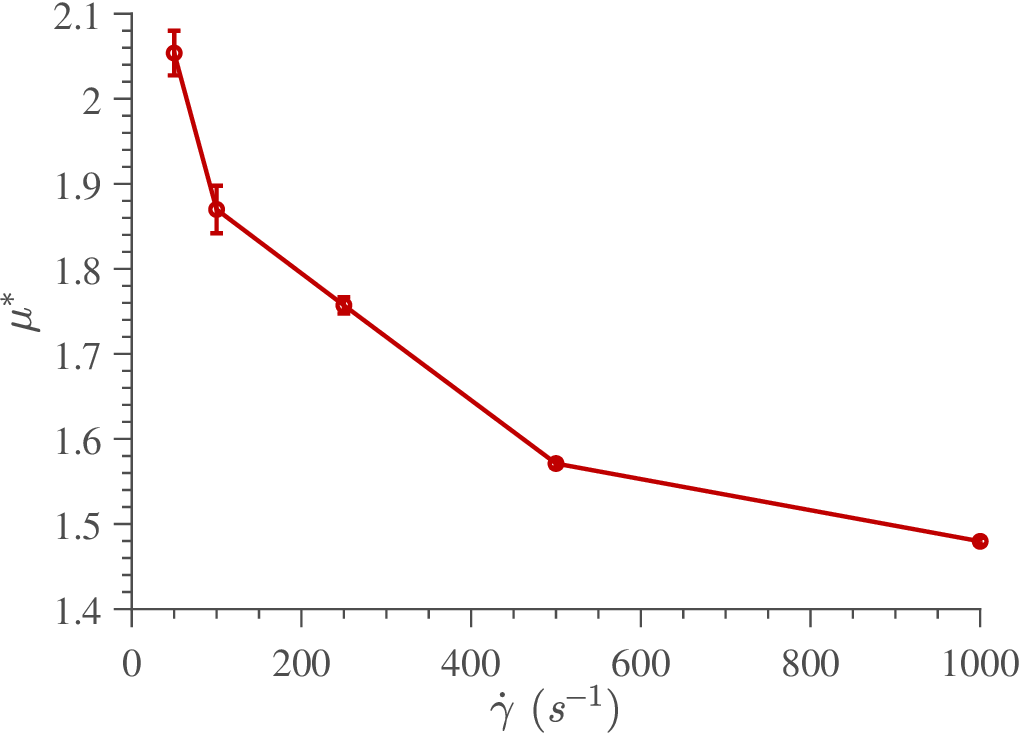}}
    \caption{Results of the effective viscosity. (a) The effective viscosity of a suspension of RBCs at a hematocrit of 25\% as a function of non-dimensional time for a shear rate of $\dot\gamma = 1000$ ${s}^{-1}$. The averaging is taken after the initial period where the {cells are initially being deformed from their resting state, at approximately 25} $\dot\gamma t$. (b) The effective viscosity of a suspension of RBCs at a hematocrit of 25\% as a function of shear rate. Note the characteristic shear-thinning behavior.}
    \label{fig:mu_eff}
\end{figure}

To select the bulk fluid viscosity for the single-cell simulations, we rely on multiple-cell simulations. More specifically, we extract the bulk viscosity from the time history of the effective viscosity that is calculated from Eq. \ref{eq:eff_visc}. An example of this time history is displayed in Fig. \ref{sub:mu_eff1} for a shear rate of $\dot \gamma = 1000$ ${s}^{-1}$. The bulk viscosity for the single-cell simulations is obtained by averaging these results after they converge. The viscosity obtained from the periodic simulations generally contains an initial period {where there is a relatively large spike as the cells initially deform} that is followed by a less regular period without sharp peaks. The averaging is taken by skipping over the initial regular period. The resulting bulk viscosity that comes out of such computations affects the Capillary number as well as the viscosity ratio $\lambda_{{v}}$ of single-cell simulations. 

The computed bulk viscosity is plotted in Fig. \ref{sub:eff_visc} as a function of shear rate. {The error bars in this figure are constructed by using the approximate spacing between peaks as a sub-averaging window. These sub-averages are used as individual points, and the error bars are the standard error of these data points.} Note that this range of shear rates is on the upper end of where the shear-thinning behavior of blood takes place. This range of shear rates is used because the problem becomes stiff at low values of the Capillary number owing to the large value of the dilatation ratio. However, the anticipated use for the solver is in the investigated range of shear rates. Regardless, the results show the characteristic shear-thinning behavior.

%

%--------------------
\subsubsection{Effect of shear rate}
Below, we compare $\lambda_1/\lambda_2$ between multiple-cell simulations using periodic boundary conditions and single-cell simulations at an equivalent effective viscosity, which is extracted from multiple-cell simulations as described in the last subsection, at varying shear rates. The multiple-cell simulations are run using parameters obtained in the previous sections, namely ${m}=12$ and $N_c=16${, as well as a hematocrit of 25\%}.

\begin{figure}[H]
    \centering
    {{\includegraphics[scale=0.35]{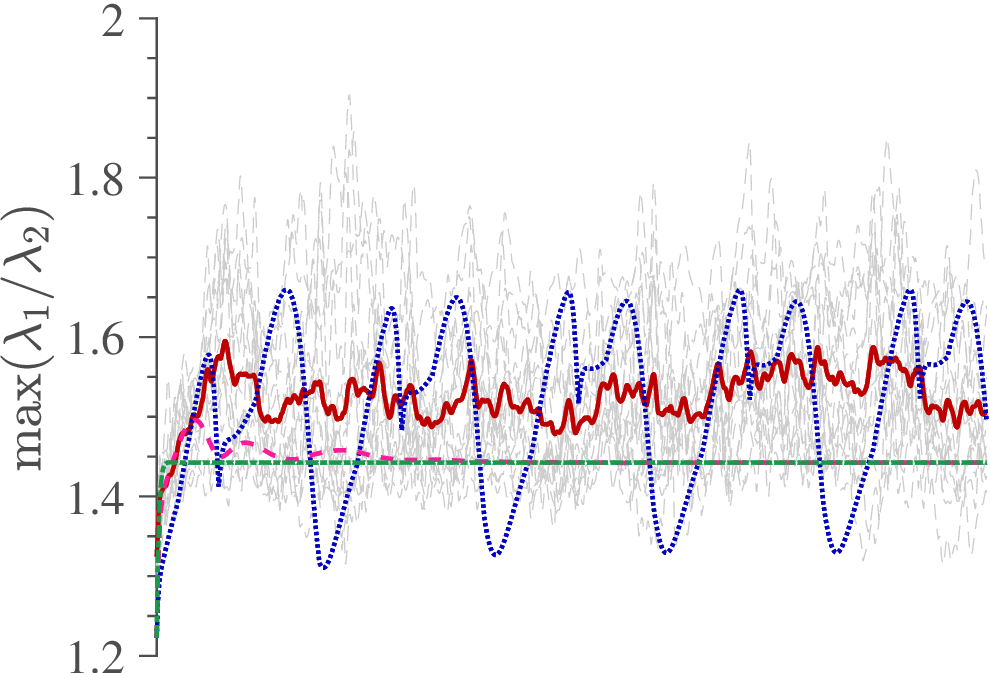}}}\\%\qquad%
    {{\includegraphics[scale=0.35]{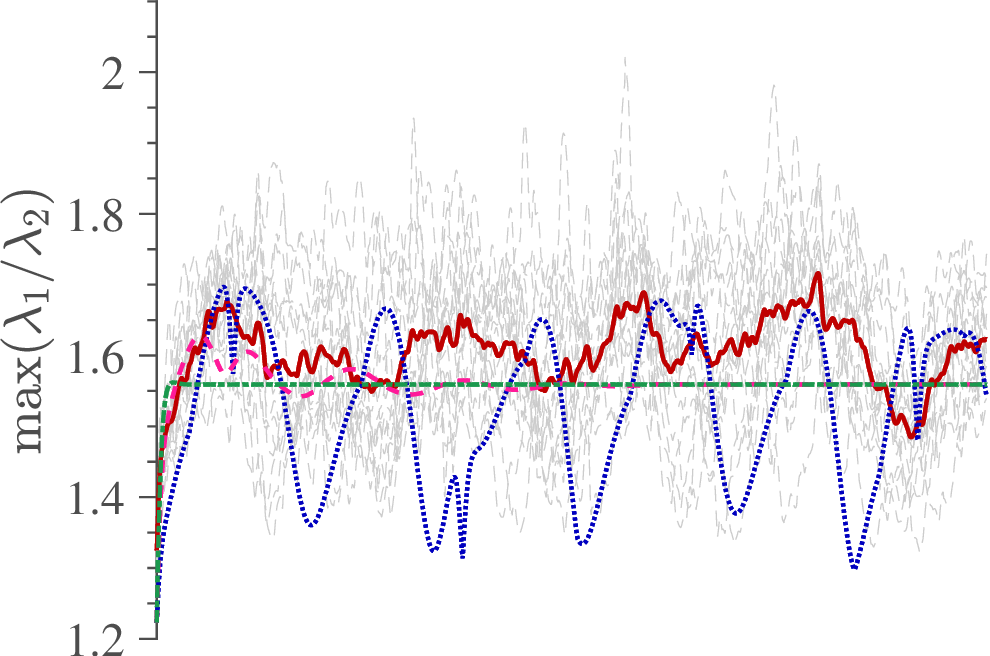}}}\\%\qquad%
    {{\includegraphics[scale=0.35]{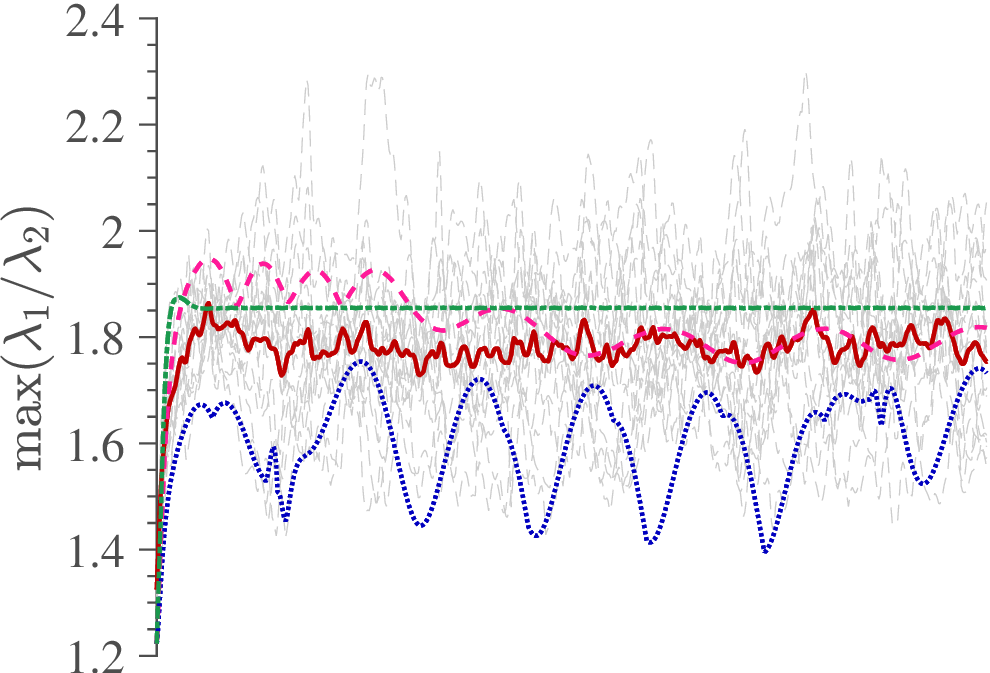}}}\\%\qquad%
    {{\includegraphics[scale=0.35]{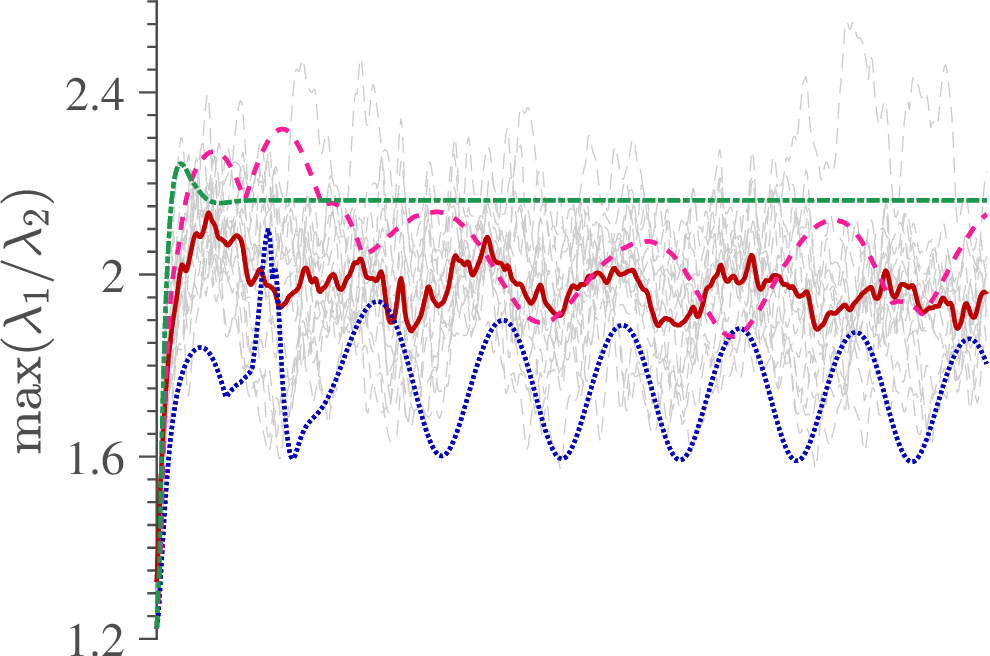}}}\\%\qquad%
    {{\includegraphics[scale=0.35]{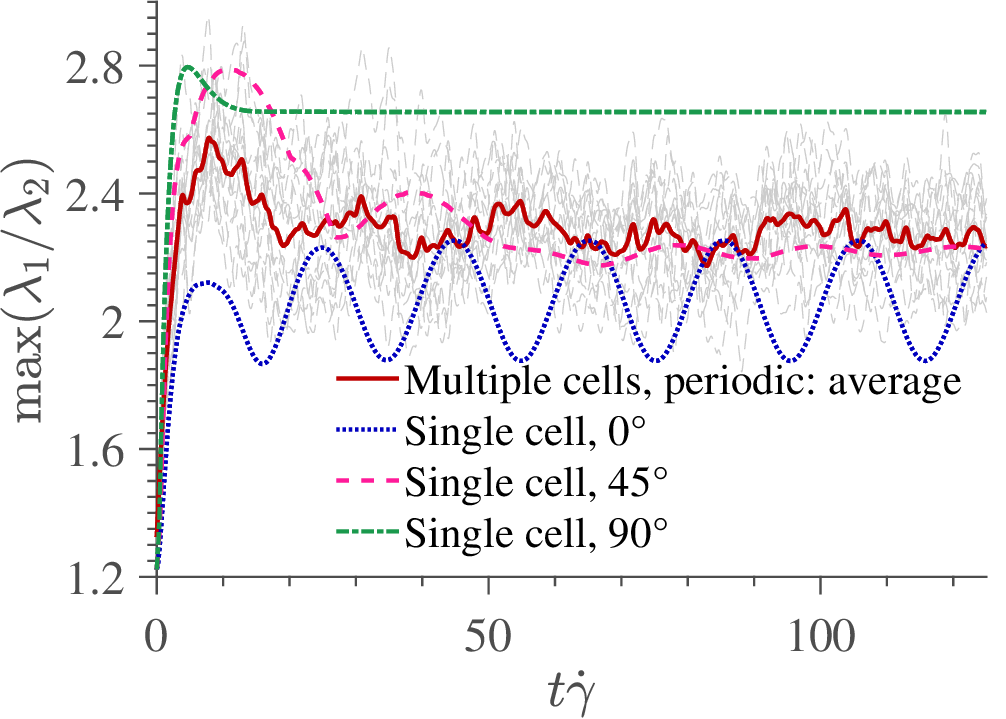}}}\\%\qquad%
    \caption{Results for the maximum value of the parameter $\lambda_1/\lambda_2$ on the surface of a cell. {From top to bottom, the results are for $\dot\gamma = 50$, $100$, $250$, $500$, and $1000$ $s^{-1}$. The results show the individual results for the multiple-cell simulations as the faint background lines, with the average of these individual results in the foreground. Additionally, the single-cell results are displayed for the three orientations shown in Fig. \ref{fig:Orientations}}.}
    \label{fig:l1l2_all}
\end{figure}

Figure \ref{fig:l1l2_all} displays the time history of $\max(\lambda_1/\lambda_2)$ at all shear rates obtained from both a cell in multiple-cell periodic simulations and a single cell at an equivalent effective viscosity, as well the spectral density of both of these signals for each case, taken after the initial start-up period. {The single-cell simulations display results for each of the three orientations shown in Fig. \ref{fig:Orientations}. The multiple-cell simulations display each of the individual results as the finer lines in the background, as well as the average of these individual cells in the foreground. The results appear to match fairly well when the individual cells and orientations are averaged, particularly at high shear rate.}

\begin{figure}[H]
   \centering
   \includegraphics[scale=.23]{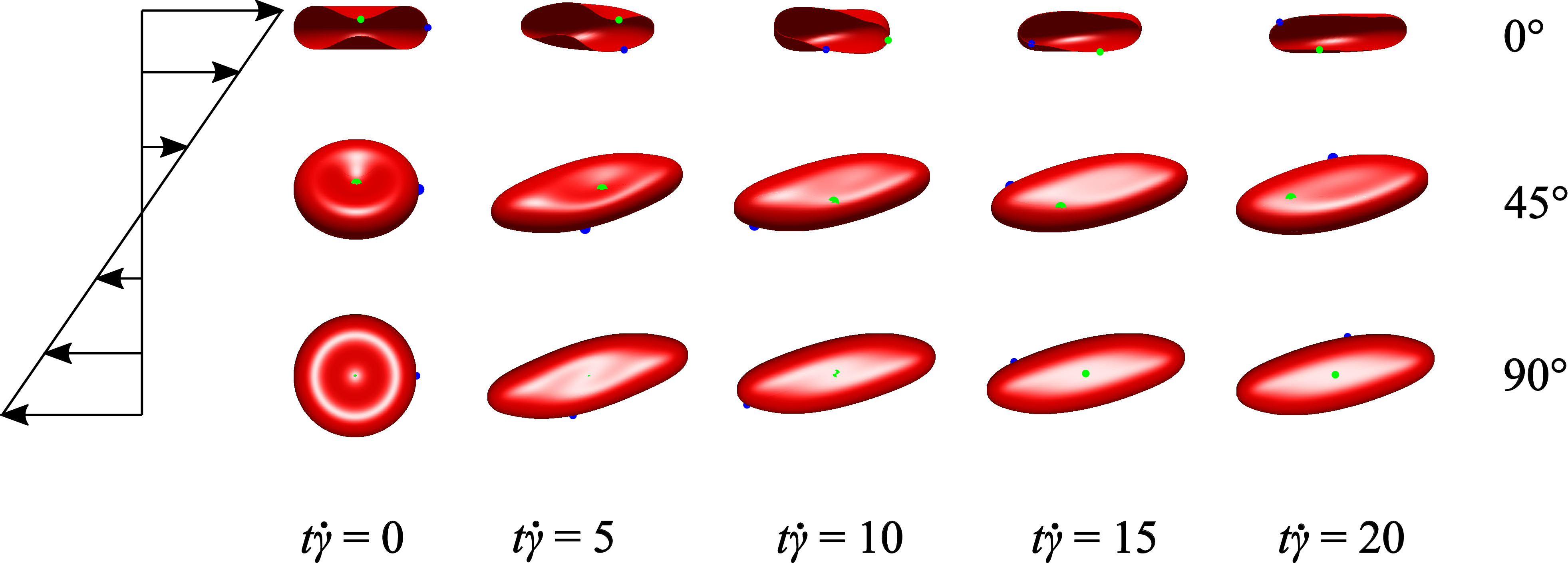}
   \caption{{Visualizations of the RBC at different points in time for the single-cell, $\dot\gamma =1000$ $s^{-1}$ simulations for the 3 orientations shown in Fig. \ref{fig:Orientations}. The two dots represent material points on the membrane surface at different instances in time.}}
   \label{fig:Visuals}
\end{figure}

{In order to visualize the single-cell simulations at different points in time, Fig. \ref{fig:Visuals} displays images of the RBCs at different points in time for each of the three orientations in Fig. \ref{fig:Orientations} at a shear rate of $\dot\gamma=1000$ $s^{-1}$.}

\begin{figure}[H]
    \centering
    \subfloat[ \label{sub:l1l2_max}] {{\includegraphics[scale=0.45]{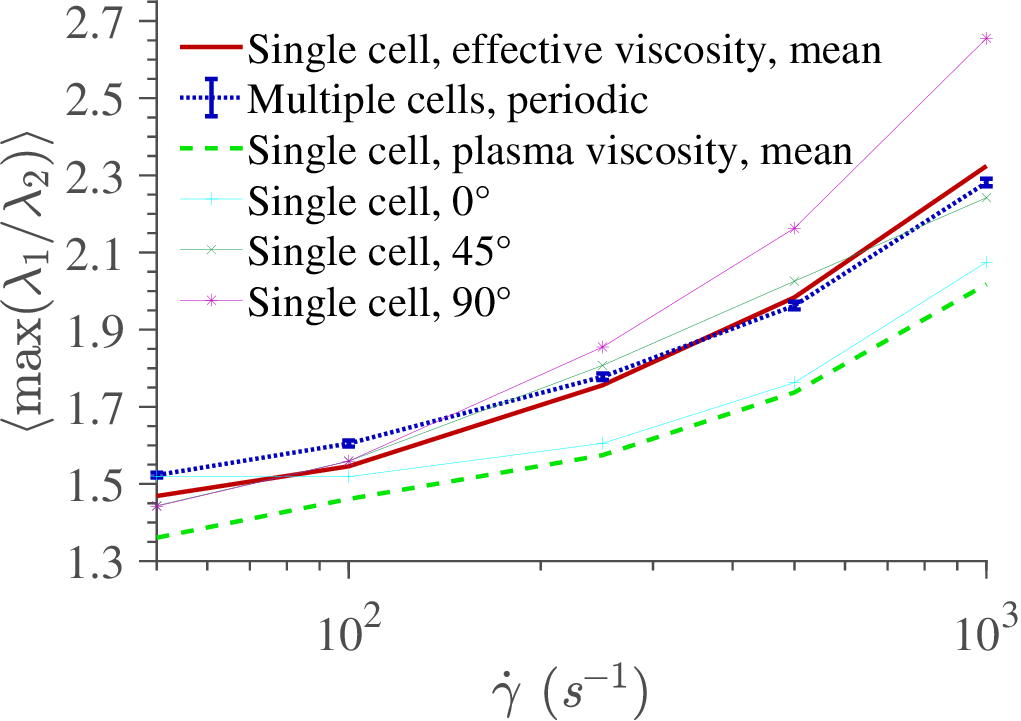}}}\quad%
    \subfloat[ \label{sub:l1l2_SR}] {\includegraphics[scale=0.45]{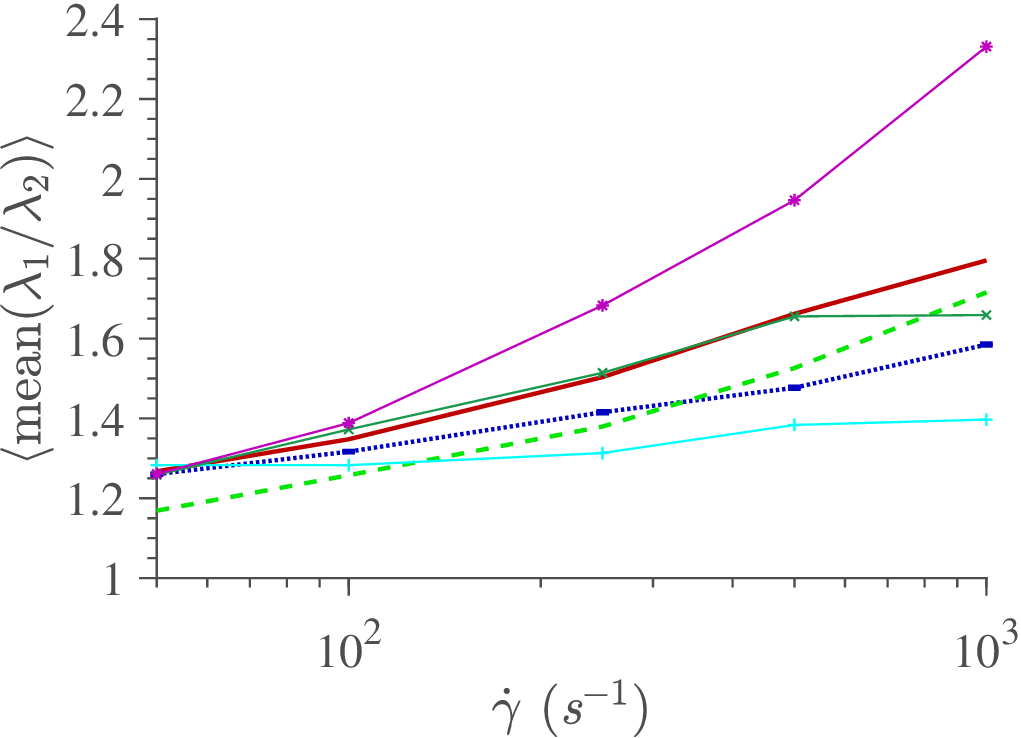}}
    \caption{{Results comparing single-cell and multiple-cell simulations as a function of shear rate. Results are plotted for both the single-cell and multiple-cell cases, averaged across all cells, as well as a single-cell case where the viscosity is not adjusted from the plasma viscosity. The single-cell, effective viscosity results are presented in terms of the three orientations in Fig. \ref{fig:Orientations} as well as the average of these orientations. To avoid clutter, the plasma viscosity cases are only presented in terms of the average of these three orientations. (a) The time-averaged value of the maximum of $\lambda_1/\lambda_2$ on the surface of the cell. (b) The time-averaged value of $\lambda_1/\lambda_2$ averaged over the surface of the cell. Good agreement is reached in the case of $\max(\lambda_1/\lambda_2)$.}}
    \label{fig:l1l2_results}
\end{figure}

Figure \ref{sub:l1l2_max} shows the time-average value of the maximum value of $\lambda_1/\lambda_2$ for both the multiple-cell periodic case and the single-cell case at an equivalent effective viscosity, as well as a single-cell case in which the viscosity is not adjusted. {For the single-cell, effective viscosity case, we have plotted the results for all orientations shown in Fig. \ref{fig:Orientations} and the average of these cases. To avoid cluttering the figures further, only the averages are plotted for the plasma viscosity case.} Note that the relatively slow increase of this value with respect to increasing shear rate is primarily a result of the strain-hardening nature of the RBC membrane. As a second measure of the similarity of the strain on the cells, Fig. \ref{sub:l1l2_SR} displays the time-average of the mean value of $\lambda_1/\lambda_2$ on the RBCs as a function of shear rate. {The error bars in these figures are constructed in the same way as those in Fig. \ref{sub:eff_visc}, except the sub-averages are taken from the results of the individual RBCs.}

% --------------------
\subsubsection{{Effect of hematocrit}}
\begin{figure}[H]
    \centering
    \subfloat[ \label{sub:hemat_time}] {{\includegraphics[scale=0.42]{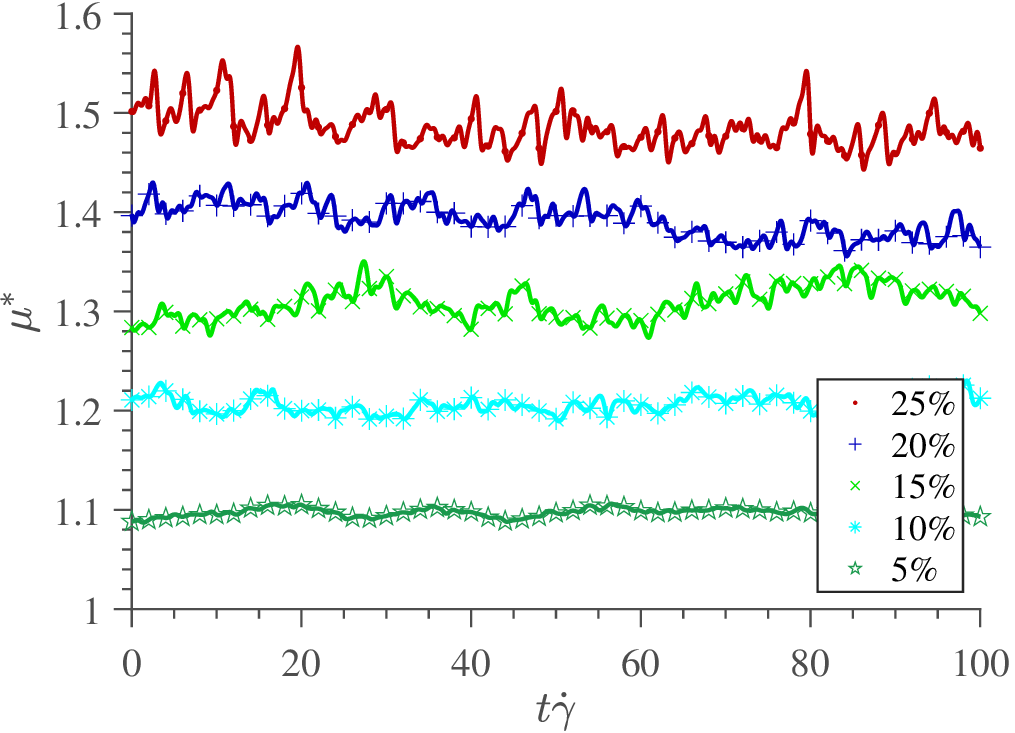}}}\qquad%
    \subfloat[ \label{sub:hemat_mean}] {\includegraphics[scale=0.42]{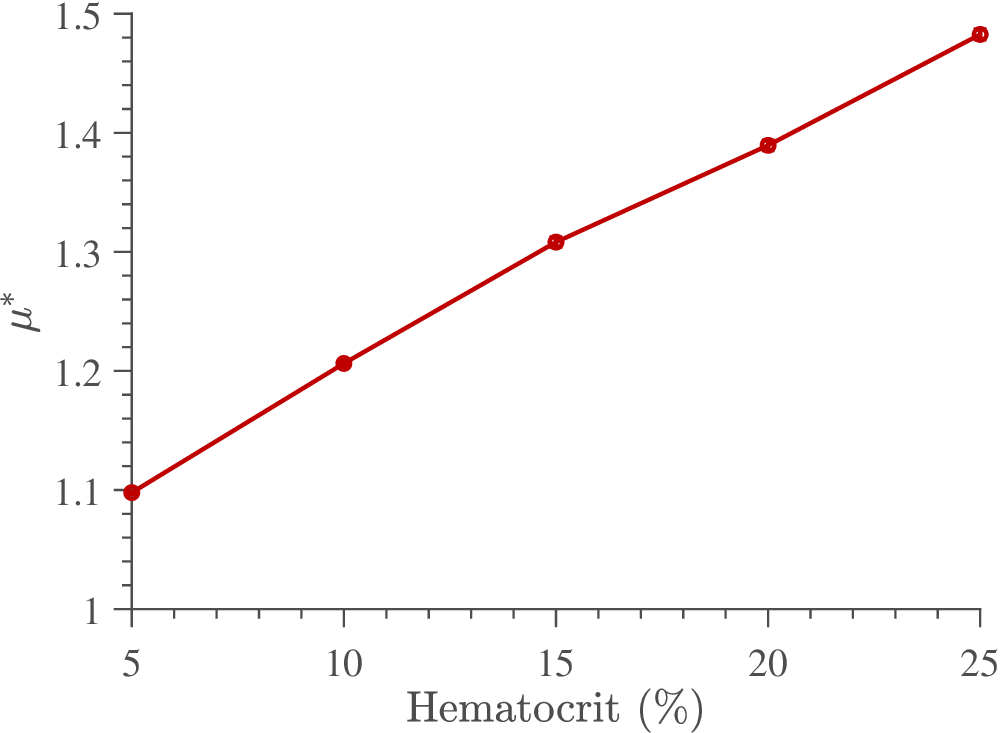}}
    \caption{{Viscosity as a function of hematocrit for simulations run at a shear rate of 1000 $s^{-1}$. (a) The time history of the viscosity as a function of hematocrit, taken after the initial starting up period, and (b) the average values of these viscosities over time.}}
    \label{fig:hemat}
\end{figure}

\begin{figure}[H]
   \centering
   \includegraphics[scale=.42]{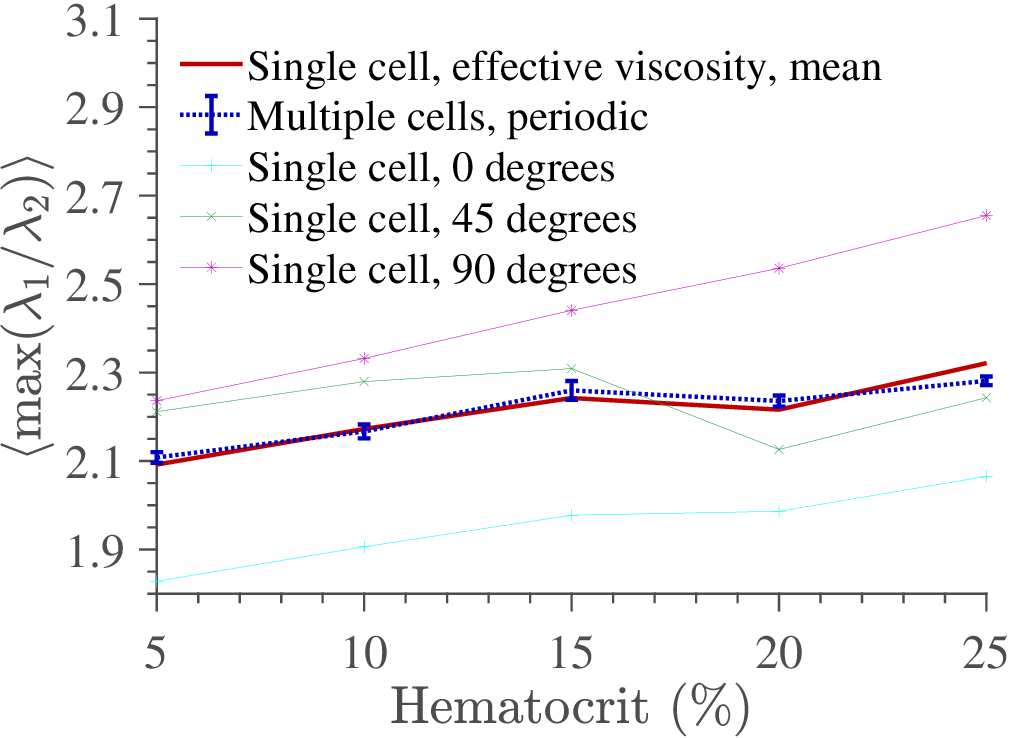}
   \caption{{The value of $\langle \max(\lambda_1/\lambda_2)\rangle$ as a function of shear rate for the multiple-cell, periodic simulations and the single-cell, effective viscosity simulations. Results for the single-cell simulations are plotted for the three orientations described in Fig. \ref{fig:Orientations}, as well as the average of these three orientations. The multiple-cell simulations match well with the average of the single-cell simulations, with a maximum difference of 1.75\%.}}
   \label{fig:hemat_l1l2}
\end{figure}

{Due to computational constraints, running simulations above a hematocrit of approximately 25\% posed some difficulties. To examine the effect of volume fraction on the results, simulations were run over a range of volume fractions from 5\% up to 25\% in order to examine the trend at these lower values of hematocrit. The below simulations are run at a shear rate of 1000 $s^{-1}$. Figure \ref{fig:hemat} displays a time history of the effective viscosities of the suspensions for a range of hematocrits, as well as the average value of these viscosities, showing a decrease in viscosity as the hematocrit decreases.}
{Figure \ref{fig:hemat_l1l2} displays the value $\langle \max(\lambda_1/\lambda_2) \rangle$ as a function of hematocrit for the multiple cell simulations and all orientations of the single-cell simulations, as well as their average. Fairly good agreement is reached, particularly at lower hematocrit, and the same trend is observed for both cases as hematocrit increases. The maximum difference between the average single-cell case and the multiple-cell case occurs at a hematocrit of 25\%, with a difference of approximately 1.75\%.}

%------------------------------
\subsubsection{Computational expense comparison}
\begin{figure}[H]
   \centering
   \includegraphics[scale=.42]{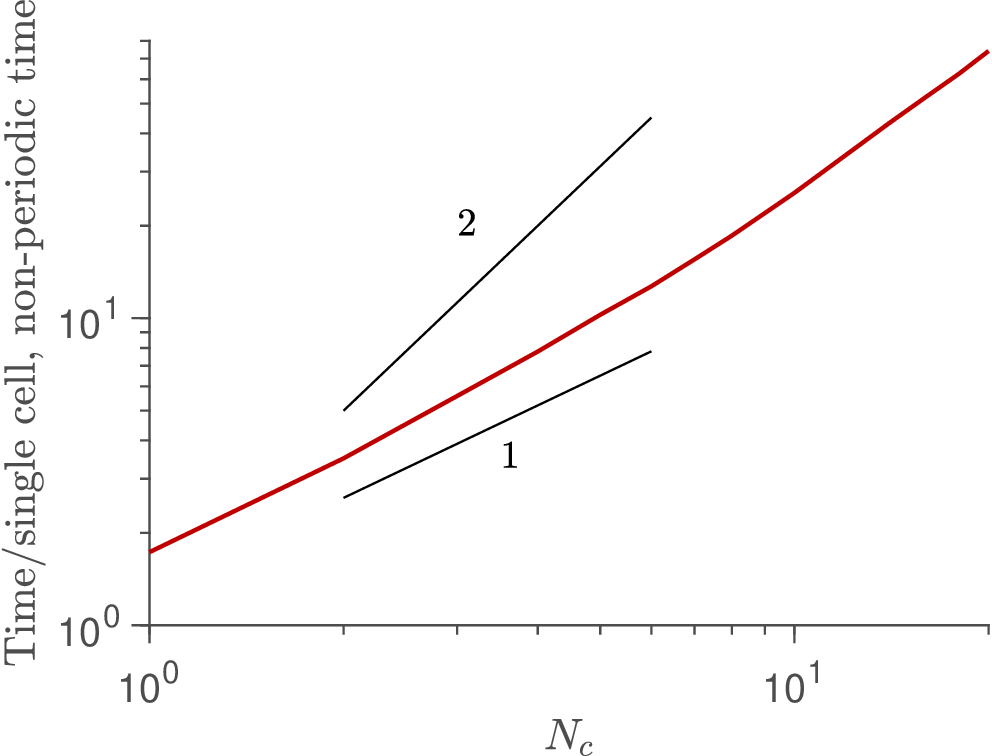}
   \caption{The average cost of the solver to run a single GMRES iteration for periodic boundary conditions as a function of the number of cells $N_c$, normalized by the cost to run an iteration for a single cell in non-periodic boundary conditions. Although our implementation is ideal in producing a nearly linear cost scaling, large cost savings can still be obtained by using single cells in place of multiple cells and periodic boundary conditions.}
   \label{fig:times}
\end{figure}

Beside its relative simplicity, a primary motivator for using single-cell simulations in lieu of multiple-cell simulations is the significant potential savings in computational cost. To express this quantitatively, the average cost of the solver per GMRES iteration is displayed as a function of the number of cells in the periodic box, normalized by the average cost per GMRES iteration of a single cell in non-periodic boundary conditions, in Fig. \ref{fig:times}. Simulations were run using a single processor. These computations are performed at a hematocrit of less than 5\% to avoid close-range interactions that would otherwise distort results in favor of single-cell simulations.  These close-range interactions in general increase the cost per iteration as they require costly near-singular integration. Furthermore, the number of GMRES iterations is fairly consistent as the number of cells increases, with the maximum per time step usually being approximately 10. Thus, we report the cost of a single GMRES iteration to evaluate the overall cost of each approach. 

The single-cell simulations present significant cost savings over the multiple-cell simulations, especially as the number of cells is increased. Note that the reported costs exclude operations that are not included in the GMRES iterations, such as geometric calculations and other pre-calculations that are carried across the GMRES iterations. These operations are typically more expensive than a single GMRES iteration, but less expensive than the sum of all the GMRES iterations. 

A naive implementation of the multiple-cell approach may easily produce a cost scaling that is $O(N_c^2)$ or worse. For instance, the tangent matrix in these computations is dense, thus filling it requires operations that scale as $O(N_c\times N_c)$ during each time step. Our matrix-free implementation circumvents this explicit matrix filling, producing a cost scaling that is approximately $O(N_c)$. Computing long-range interactions of the cells is another part of the implementation that, if not done carefully, can produce $O(N_c^2)$ scaling. The fast Ewald methods \cite{af_klinteberg_fast_2014} adopted here avoid this issue, leading to the ideal linear cost scaling with respect to $N_c$ (Fig. \ref{fig:times}). 

%------------------------------
\section{Discussion}\label{sec:discussion}
In the above simulations, particularly notable in Figs. \ref{sub:l1l2_max} and {\ref{fig:hemat_l1l2}}, the single-cell simulations satisfactorily track the behavior of the multiple-cell simulations with respect to {the maximum strain on the cell surface. However, the single-cell simulations typically over-predict the average strain on the cell. This appears to be primarily due to the 90\textdegree\, orientation of Fig. \ref{fig:Orientations}, which produces elevated levels of strain relative to the multiple-cell simulations.} Across the range of shear rates tested, the maximum difference in magnitude of the time-averaged values of {$\max(\lambda_1/\lambda_2)$ is 3.7\%, occurring at $\dot\gamma = 100$ $s^{-1}$. There is less agreement on the mean value of $\lambda_1/\lambda_2$, however, with a maximum difference of 13\% at $\dot\gamma = 1000$ $s^{-1}$. These differences appear to get worse as the shear rate increases. Likely the most important parameter with respect to hemolysis is the maximum strain, which matches well across all cases, as these areas of maximum strain are likely where hemolysis would occur. This trend persists across a range of hematocrit cases when $\dot\gamma$ was held constant at $1000$ $s^{-1}$ as well. In fact, the maximum difference between the average of the single-cell cases and the multiple-cell case was 1.9\% in Fig. \ref{fig:hemat_l1l2}.} 

{These results are of high significance, as the single-cell simulations are far more cost-effective and less elaborate to implement, set up, and perform. Despite all optimizations utilized in developing the present computational method, we observe that the multiple-cell simulations with $N_c=16$ adopted in this study are at least 50 times more expensive than those of the single-cell simulations (Fig. \ref{fig:times}). In more extreme cases at 25\% hematocrit, this cost disparity can be as high as nearly 67 times. Therefore, the ability to use single-cell simulations while faithfully capturing the RBC mechanics presents a significant cost advantage, making its adoption more practical for engineering applications.}

The simulations in which the viscosity is not adjusted highlight the importance of using the correct bulk viscosity for single-cell simulations. Across {shear rates}, the bulk viscosity simulations perform better than the plasma viscosity cases {with respect to the value of $\max(\lambda_1/\lambda_2)$. The minimum difference between the plasma viscosity case and the multiple-cell case is 5.3\% at $\dot\gamma = 100$ $s^{-1}$ and a maximum difference of 13\% at $\dot\gamma = 1000$ $s^{-1}$. Thus, by simply adjusting the fluid viscosity outside of the RBC to that of the whole blood, which is known experimentally for a wide range of conditions, one can produce much more accurate predictions using single-cell simulations.}

{The visualizations in Fig. \ref{fig:Visuals} explain some of the features of the single cell simulations in Fig. \ref{fig:l1l2_all}, as well as, perhaps, some of the differences between the cases in Fig. \ref{fig:l1l2_results}. The 0\textdegree\, simulations undergo less relative extension at this shear rate, leading to smaller values of $\lambda_1/\lambda_2$ overall. Additionally, this case undergoes periodic deformation, as the shape of the membrane changes slightly as the cell tank-treads. In contrast, the 90\textdegree\, case retains a constant shape due to the symmetry of the cell relative to the shear flow. For example, note that the central material point does not shift positions. The 45\textdegree\, case, on the other hand, settles into a somewhat stable shape with minor oscillations. The fact that these orientations are weighted equally could be responsible for some of the discrepancies observed between single- and multiple-cell simulations. For example, examining Fig. \ref{fig:RBC_side}, the majority of cells appear to be oriented approximately between the 0\textdegree\, and 45\textdegree\, cases, although calculating this orientation is not straightforward due to the tank-treading motion of the cells.}

% --------------------
\subsection{{Limitations}}\label{sec:limitations}
{Firstly, the comparison between single- and multiple-cell simulation was limited to shear flows. The validity of single-cell simulations in an extensional flow remains an open question. Secondly, the present study neglected the effect of higher-order spatial variations in the far-field velocity on the results. This assumption, however, is well justified given that the RBC is much smaller than the viscous length scale in physiological flows. Incorporating such effects, although straightforward in non-periodic simulations in unbounded flows, is not trivial for periodic calculations.} {Thirdly, the cells are modeled as two-dimensional surfaces, and many microscopic phenomena are not captured by such a model. Fourthly, since close-range attractive forces are neglected, the current model may not be suitable for low shear rates where these forces are important. Fifthly, the accurate application of the single-cell simulation framework relies on the availability of viscosity data. Ideally, viscosity data could be taken from experiments or models. However, experimental data or applicable models may not always be available. In such a case, viscosities would likely need to be calculated across a range of parameters of interest, which would be computationally expensive.} {Lastly, the present framework for modeling RBCs ought not to be used by itself in macroscale flows where inertial effects are important. The intended use of this technology for such cases is in conjunction with a CFD solver, where the velocity gradient along cell trajectories is extracted from the macroscale flow (following completion of a CFD simulation) and imposed as the far-field velocity for the present computations in a post-processing step. }

% --------------------
\subsection{Future work}\label{sec:future}
Future steps involve using the solver in realistic scenarios. The differences between the single-cell and multiple-cell simulations suggest that a modified effective viscosity is adequate at representing the maximum strain stemming from the cell-cell interactions. This reduces the cost of running many simulations corresponding to a large number of pathlines, thus allowing for an economical sampling of geometries of interest.

Additionally, a model to represent damage on the cells should be implemented. Currently, the solver can only evaluate potential geometries in a relative sense; the solver can evaluate the strains experienced by cells in different geometries and use that as an indicator of potential damage, but it cannot be used to evaluate the expected level of hemolysis yet. Implementing an explicit damage model would likely involve implementing, e.g., pore formation and hemoglobin leakage, and could be built on top of the solver in its current state.
% --------------------
\section{Conclusions}
We investigated whether simulations involving a single RBC with an effective fluid viscosity are accurate in reproducing those with multiple interacting RBCs. This question was investigated for RBCs in shear flow using a boundary integral solver to resolve the dynamics of the cells. Periodic boundary conditions are used to replicate the long-range interactions of the cells. We showed that a periodic box size that contains {8} cells is sufficiently large to produce converged results. Similarly, we found the lowest degree for the spherical harmonic to produce converged results is {10}. The results showed that the single-cell simulations replicate the multiple-cell simulations in terms of {maximum} strain over the cell (maximum difference was {3.7\%} over the studied range of shear rates, occurring at a shear rate of {$\dot\gamma=100$ $s^{-1}$}). Additionally, the simulations in which the viscosity is adjusted to the bulk viscosity unilaterally perform better than those in which the plasma viscosity is used {when comparing the maximum value of $\lambda_1/\lambda_2$. This trend persists across hematocrit as well at a shear rate of $\dot\gamma = 1000$ $s^{-1}$}. {In conclusion, although single-cell simulations present several orders of magnitude cost advantage over multiple-cell simulations, one must be careful in its adoption to appropriately adjust the bulk viscosity and average over multiple cell orientations when analyzing its predictions. }
% --------------------
\section*{Acknowledgements}
Research reported in this publication was supported by the National Heart, Lung, and Blood Institute of the National Institutes of Health under award number R01HL089456-10.

% --------------------
\section*{Conflict of Interest}
The authors declare no conflict of interest.
 
%--------------------
% REFERENCES SECTION
\color{black}
\medskip
\bibliographystyle{elsarticle-num}
\bibliography{references} 

\begin{thebibliography}{10}
\expandafter\ifx\csname url\endcsname\relax
  \def\url#1{\texttt{#1}}\fi
\expandafter\ifx\csname urlprefix\endcsname\relax\def\urlprefix{URL }\fi
\expandafter\ifx\csname href\endcsname\relax
  \def\href#1#2{#2} \def\path#1{#1}\fi

\bibitem{yu_review_2017}
H.~Yu, S.~Engel, G.~Janiga, D.~Thévenin, A review of hemolysis prediction models for computational fluid dynamics, Artificial Organs 41~(7) (2017) 603--621.
\newblock \href {http://dx.doi.org/10.1111/aor.12871} {\path{doi:10.1111/aor.12871}}.

\bibitem{giersiepen_estimation_1990}
M.~Giersiepen, L.~J. Wurzinger, R.~Opitz, H.~Reul, Estimation of shear stress-related blood damage in heart valve prostheses--in vitro comparison of 25 aortic valves, The International Journal of Artificial Organs 13~(5) (1990-05) 300--306.

\bibitem{sharp_scaling_1998}
M.~K. Sharp, S.~F. Mohammad, \href{https://doi.org/10.1114/1.65}{Scaling of hemolysis in needles and catheters}, Annals of Biomedical Engineering 26~(5) (1998-09-01) 788--797.
\newblock \href {http://dx.doi.org/10.1114/1.65} {\path{doi:10.1114/1.65}}.
\newline\urlprefix\url{https://doi.org/10.1114/1.65}

\bibitem{arora_tensor-based_2004}
D.~Arora, M.~Behr, M.~Pasquali, \href{http://onlinelibrary.wiley.com/doi/abs/10.1111/j.1525-1594.2004.00072.x}{A tensor-based measure for estimating blood damage}, Artificial Organs 28~(11) (2004) 1002--1015.
\newblock \href {http://dx.doi.org/https://doi.org/10.1111/j.1525-1594.2004.00072.x} {\path{doi:https://doi.org/10.1111/j.1525-1594.2004.00072.x}}.
\newline\urlprefix\url{http://onlinelibrary.wiley.com/doi/abs/10.1111/j.1525-1594.2004.00072.x}

\bibitem{arvand_validated_2005}
A.~Arvand, M.~Hormes, H.~Reul, \href{http://onlinelibrary.wiley.com/doi/abs/10.1111/j.1525-1594.2005.29089.x}{A validated computational fluid dynamics model to estimate hemolysis in a rotary blood pump}, Artificial Organs 29~(7) (2005) 531--540.
\newblock \href {http://dx.doi.org/10.1111/j.1525-1594.2005.29089.x} {\path{doi:10.1111/j.1525-1594.2005.29089.x}}.
\newline\urlprefix\url{http://onlinelibrary.wiley.com/doi/abs/10.1111/j.1525-1594.2005.29089.x}

\bibitem{ezzeldin_strain-based_2015}
H.~M. Ezzeldin, M.~D. de~Tullio, M.~Vanella, S.~D. Solares, E.~Balaras, \href{https://doi.org/10.1007/s10439-015-1273-z}{A strain-based model for mechanical hemolysis based on a coarse-grained red blood cell model}, Annals of Biomedical Engineering 43~(6) (2015-06-01) 1398--1409.
\newblock \href {http://dx.doi.org/10.1007/s10439-015-1273-z} {\path{doi:10.1007/s10439-015-1273-z}}.
\newline\urlprefix\url{https://doi.org/10.1007/s10439-015-1273-z}

\bibitem{sohrabi_cellular_2017}
S.~Sohrabi, Y.~Liu, \href{http://onlinelibrary.wiley.com/doi/abs/10.1111/aor.12832}{A cellular model of shear-induced hemolysis}, Artificial Organs 41~(9) (2017) E80--E91.
\newblock \href {http://dx.doi.org/https://doi.org/10.1111/aor.12832} {\path{doi:https://doi.org/10.1111/aor.12832}}.
\newline\urlprefix\url{http://onlinelibrary.wiley.com/doi/abs/10.1111/aor.12832}

\bibitem{nikfar_multiscale_2020}
M.~Nikfar, M.~Razizadeh, R.~Paul, Y.~Liu, \href{https://doi.org/10.1007/s10404-020-02337-3}{Multiscale modeling of hemolysis during microfiltration}, Microfluidics and Nanofluidics 24~(5) (2020-04-10) 33.
\newblock \href {http://dx.doi.org/10.1007/s10404-020-02337-3} {\path{doi:10.1007/s10404-020-02337-3}}.
\newline\urlprefix\url{https://doi.org/10.1007/s10404-020-02337-3}

\bibitem{garon_fast_2004}
A.~Garon, M.-I. Farinas, Fast three-dimensional numerical hemolysis approximation, Artificial Organs 28~(11) (2004) 1016--1025.
\newblock \href {http://dx.doi.org/10.1111/j.1525-1594.2004.00026.x} {\path{doi:10.1111/j.1525-1594.2004.00026.x}}.

\bibitem{zhang_computational_2006}
J.~Zhang, B.~Gellman, A.~Koert, K.~A. Dasse, R.~J. Gilbert, B.~P. Griffith, Z.~J. Wu, \href{https://onlinelibrary.wiley.com/doi/abs/10.1111/j.1525-1594.2006.00203.x}{Computational and {Experimental} {Evaluation} of the {Fluid} {Dynamics} and {Hemocompatibility} of the {CentriMag} {Blood} {Pump}}, Artificial Organs 30~(3) (2006) 168--177.
\newblock \href {http://dx.doi.org/10.1111/j.1525-1594.2006.00203.x} {\path{doi:10.1111/j.1525-1594.2006.00203.x}}.
\newline\urlprefix\url{https://onlinelibrary.wiley.com/doi/abs/10.1111/j.1525-1594.2006.00203.x}

\bibitem{grigioni_power-law_2004}
M.~Grigioni, C.~Daniele, U.~Morbiducci, G.~D'Avenio, G.~Di~Benedetto, V.~Barbaro, The power-law mathematical model for blood damage prediction: analytical developments and physical inconsistencies, Artificial Organs 28~(5) (2004) 467--475.
\newblock \href {http://dx.doi.org/10.1111/j.1525-1594.2004.00015.x} {\path{doi:10.1111/j.1525-1594.2004.00015.x}}.

\bibitem{goubergrits_numerical_2004}
L.~Goubergrits, K.~Affeld, Numerical estimation of blood damage in artificial organs, Artificial Organs 28~(5) (2004) 499--507.
\newblock \href {http://dx.doi.org/10.1111/j.1525-1594.2004.07265.x} {\path{doi:10.1111/j.1525-1594.2004.07265.x}}.

\bibitem{tamagawa_prediction_1996}
M.~Tamagawa, T.~Akamatsu, K.~Saitoh, \href{https://onlinelibrary.wiley.com/doi/abs/10.1111/j.1525-1594.1996.tb04479.x}{Prediction of {Hemolysis} in {Turbulent} {Shear} {Orifice} {Flow}}, Artificial Organs 20~(5) (1996) 553--559.
\newblock \href {http://dx.doi.org/10.1111/j.1525-1594.1996.tb04479.x} {\path{doi:10.1111/j.1525-1594.1996.tb04479.x}}.
\newline\urlprefix\url{https://onlinelibrary.wiley.com/doi/abs/10.1111/j.1525-1594.1996.tb04479.x}

\bibitem{ozturk_approach_2016}
M.~Ozturk, D.~V. Papavassiliou, E.~A. O'Rear, \href{https://doi.org/10.1115/1.4034992}{An {Approach} for {Assessing} {Turbulent} {Flow} {Damage} to {Blood} in {Medical} {Devices}}, Journal of Biomechanical Engineering 139~(011008).
\newblock \href {http://dx.doi.org/10.1115/1.4034992} {\path{doi:10.1115/1.4034992}}.
\newline\urlprefix\url{https://doi.org/10.1115/1.4034992}

\bibitem{goubergrits_turbulence_2016}
L.~Goubergrits, J.~Osman, R.~Mevert, U.~Kertzscher, K.~Pöthkow, H.-C. Hege, Turbulence in blood damage modeling, The International Journal of Artificial Organs 39~(4) (2016) 160--165.
\newblock \href {http://dx.doi.org/10.5301/ijao.5000476} {\path{doi:10.5301/ijao.5000476}}.

\bibitem{balogh_direct_2017}
P.~Balogh, P.~Bagchi, \href{https://linkinghub.elsevier.com/retrieve/pii/S0006349517311359}{Direct {Numerical} {Simulation} of {Cellular}-{Scale} {Blood} {Flow} in {3D} {Microvascular} {Networks}}, Biophysical Journal 113~(12) (2017) 2815--2826.
\newblock \href {http://dx.doi.org/10.1016/j.bpj.2017.10.020} {\path{doi:10.1016/j.bpj.2017.10.020}}.
\newline\urlprefix\url{https://linkinghub.elsevier.com/retrieve/pii/S0006349517311359}

\bibitem{lu_scalable_2019}
L.~Lu, M.~J. Morse, A.~Rahimian, G.~Stadler, D.~Zorin, \href{http://doi.org/10.1145/3295500.3356203}{Scalable simulation of realistic volume fraction red blood cell flows through vascular networks}, in: Proceedings of the {International} {Conference} for {High} {Performance} {Computing}, {Networking}, {Storage} and {Analysis}, {SC} '19, Association for Computing Machinery, New York, NY, USA, 2019, pp. 1--30.
\newblock \href {http://dx.doi.org/10.1145/3295500.3356203} {\path{doi:10.1145/3295500.3356203}}.
\newline\urlprefix\url{http://doi.org/10.1145/3295500.3356203}

\bibitem{peters_multiscale_2010}
A.~Peters, S.~Melchionna, E.~Kaxiras, J.~Lätt, J.~Sircar, M.~Bernaschi, M.~Bison, S.~Succi, Multiscale {Simulation} of {Cardiovascular} flows on the {IBM} {Bluegene}/{P}: {Full} {Heart}-{Circulation} {System} at {Red}-{Blood} {Cell} {Resolution}, in: {SC} '10: {Proceedings} of the 2010 {ACM}/{IEEE} {International} {Conference} for {High} {Performance} {Computing}, {Networking}, {Storage} and {Analysis}, 2010, pp. 1--10, iSSN: 2167-4337.
\newblock \href {http://dx.doi.org/10.1109/SC.2010.33} {\path{doi:10.1109/SC.2010.33}}.

\bibitem{rydquist_cell-resolved_2022}
G.~Rydquist, M.~Esmaily, \href{https://www.sciencedirect.com/science/article/pii/S0021999122002662}{A cell-resolved, {Lagrangian} solver for modeling red blood cell dynamics in macroscale flows}, Journal of Computational Physics 461 (2022) 111204.
\newblock \href {http://dx.doi.org/10.1016/j.jcp.2022.111204} {\path{doi:10.1016/j.jcp.2022.111204}}.
\newline\urlprefix\url{https://www.sciencedirect.com/science/article/pii/S0021999122002662}

\bibitem{faghih_modeling_2019}
M.~M. Faghih, M.~K. Sharp, \href{https://doi.org/10.1007/s10237-019-01137-1}{Modeling and prediction of flow-induced hemolysis: a review}, Biomechanics and Modeling in Mechanobiology 18~(4) (2019) 845--881.
\newblock \href {http://dx.doi.org/10.1007/s10237-019-01137-1} {\path{doi:10.1007/s10237-019-01137-1}}.
\newline\urlprefix\url{https://doi.org/10.1007/s10237-019-01137-1}

\bibitem{chien_shear_1970}
S.~Chien, \href{http://science.sciencemag.org/content/168/3934/977}{Shear {Dependence} of {Effective} {Cell} {Volume} as a {Determinant} of {Blood} {Viscosity}}, Science 168~(3934) (1970) 977--979, publisher: American Association for the Advancement of Science Section: Reports.
\newblock \href {http://dx.doi.org/10.1126/science.168.3934.977} {\path{doi:10.1126/science.168.3934.977}}.
\newline\urlprefix\url{http://science.sciencemag.org/content/168/3934/977}

\bibitem{fedosov_predicting_2011}
D.~A. Fedosov, W.~Pan, B.~Caswell, G.~Gompper, G.~E. Karniadakis, \href{http://www.pnas.org/content/108/29/11772}{Predicting human blood viscosity in silico}, Proceedings of the National Academy of Sciences 108~(29) (2011) 11772--11777, publisher: National Academy of Sciences Section: Physical Sciences.
\newblock \href {http://dx.doi.org/10.1073/pnas.1101210108} {\path{doi:10.1073/pnas.1101210108}}.
\newline\urlprefix\url{http://www.pnas.org/content/108/29/11772}

\bibitem{leverett_red_1972}
L.~B. Leverett, J.~D. Hellums, C.~P. Alfrey, E.~C. Lynch, Red blood cell damage by shear stress, Biophysical Journal 12~(3) (1972) 257--273.
\newblock \href {http://dx.doi.org/10.1016/S0006-3495(72)86085-5} {\path{doi:10.1016/S0006-3495(72)86085-5}}.

\bibitem{porcaro_hemolysis_2023}
C.~Porcaro, M.~Saeedipour, Hemolysis prediction in bio-microfluidic applications using resolved {CFD}-{DEM} simulations, Computer Methods and Programs in Biomedicine 231 (2023) 107400.
\newblock \href {http://dx.doi.org/10.1016/j.cmpb.2023.107400} {\path{doi:10.1016/j.cmpb.2023.107400}}.

\bibitem{skalak_strain_1973}
R.~Skalak, A.~Tozeren, R.~P. Zarda, S.~Chien, \href{https://www.ncbi.nlm.nih.gov/pmc/articles/PMC1484188/}{Strain {Energy} {Function} of {Red} {Blood} {Cell} {Membranes}}, Biophysical Journal 13~(3) (1973) 245--264.
\newline\urlprefix\url{https://www.ncbi.nlm.nih.gov/pmc/articles/PMC1484188/}

\bibitem{helfrich_elastic_1973}
W.~Helfrich, Elastic {Properties} of {Lipid} {Bilayers}: {Theory} and {Possible} {Experiments}, Z. Naturforsch. C 28.
\newblock \href {http://dx.doi.org/10.1515/znc-1973-11-1209} {\path{doi:10.1515/znc-1973-11-1209}}.

\bibitem{barthes-biesel_effect_2002}
D.~Barthès-Biesel, A.~Diaz, E.~Dhenin, Effect of constitutive laws for two-dimensional membranes on flow-induced capsule deformation, Journal of Fluid Mechanics 460 (2002) 211--222, publisher: Cambridge University Press.
\newblock \href {http://dx.doi.org/10.1017/S0022112002008352} {\path{doi:10.1017/S0022112002008352}}.

\bibitem{pozrikidis_numerical_2003}
C.~Pozrikidis, \href{https://doi.org/10.1114/1.1617985}{Numerical {Simulation} of the {Flow}-{Induced} {Deformation} of {Red} {Blood} {Cells}}, Annals of Biomedical Engineering 31~(10) (2003) 1194--1205.
\newblock \href {http://dx.doi.org/10.1114/1.1617985} {\path{doi:10.1114/1.1617985}}.
\newline\urlprefix\url{https://doi.org/10.1114/1.1617985}

\bibitem{zhao_spectral_2010}
H.~Zhao, A.~H. Isfahani, L.~N. Olson, J.~B. Freund, \href{https://linkinghub.elsevier.com/retrieve/pii/S0021999110000471}{A spectral boundary integral method for flowing blood cells}, Journal of Computational Physics 229~(10) (2010) 3726--3744.
\newblock \href {http://dx.doi.org/10.1016/j.jcp.2010.01.024} {\path{doi:10.1016/j.jcp.2010.01.024}}.
\newline\urlprefix\url{https://linkinghub.elsevier.com/retrieve/pii/S0021999110000471}

\bibitem{sinha_dynamics_2015}
K.~Sinha, M.~D. Graham, \href{https://link.aps.org/doi/10.1103/PhysRevE.92.042710}{Dynamics of a single red blood cell in simple shear flow}, Physical Review E 92~(4) (2015) 042710, publisher: American Physical Society.
\newblock \href {http://dx.doi.org/10.1103/PhysRevE.92.042710} {\path{doi:10.1103/PhysRevE.92.042710}}.
\newline\urlprefix\url{https://link.aps.org/doi/10.1103/PhysRevE.92.042710}

\bibitem{pozrikidis_boundary_1992}
C.~Pozrikidis, \href{https://www.cambridge.org/core/books/boundary-integral-and-singularity-methods-for-linearized-viscous-flow/219AA1CE5DE05AE67096EA59692E25B3}{Boundary Integral and Singularity Methods for Linearized Viscous Flow}, Cambridge Texts in Applied Mathematics, Cambridge University Press, 1992.
\newblock \href {http://dx.doi.org/10.1017/CBO9780511624124} {\path{doi:10.1017/CBO9780511624124}}.
\newline\urlprefix\url{https://www.cambridge.org/core/books/boundary-integral-and-singularity-methods-for-linearized-viscous-flow/219AA1CE5DE05AE67096EA59692E25B3}

\bibitem{pozrikidis_computation_1996}
C.~Pozrikidis, \href{https://doi.org/10.1007/BF00118824}{Computation of periodic {Green}'s functions of {Stokes} flow}, Journal of Engineering Mathematics 30~(1) (1996) 79--96.
\newblock \href {http://dx.doi.org/10.1007/BF00118824} {\path{doi:10.1007/BF00118824}}.
\newline\urlprefix\url{https://doi.org/10.1007/BF00118824}

\bibitem{lindbo_spectrally_2010}
D.~Lindbo, A.-K. Tornberg, \href{https://www.sciencedirect.com/science/article/pii/S0021999110004730}{Spectrally accurate fast summation for periodic {Stokes} potentials}, Journal of Computational Physics 229~(23) (2010) 8994--9010.
\newblock \href {http://dx.doi.org/10.1016/j.jcp.2010.08.026} {\path{doi:10.1016/j.jcp.2010.08.026}}.
\newline\urlprefix\url{https://www.sciencedirect.com/science/article/pii/S0021999110004730}

\bibitem{marin_boundary_2012}
O.~Marin, Boundary integral methods for {Stokes} flow : {Quadrature} techniques and fast {Ewald} methods, Ph.D. thesis, KTH Royal Institute of Technology (2012).

\bibitem{af_klinteberg_fast_2014}
L.~Af~Klinteberg, A.-K. Tornberg, \href{http://onlinelibrary.wiley.com/doi/abs/10.1002/fld.3953}{Fast {Ewald} summation for {Stokesian} particle suspensions}, International Journal for Numerical Methods in Fluids 76~(10) (2014) 669--698.
\newblock \href {http://dx.doi.org/10.1002/fld.3953} {\path{doi:10.1002/fld.3953}}.
\newline\urlprefix\url{http://onlinelibrary.wiley.com/doi/abs/10.1002/fld.3953}

\bibitem{af_klinteberg_fast_2017}
L.~af~Klinteberg, D.~S. Shamshirgar, A.-K. Tornberg, \href{https://doi.org/10.1186/s40687-016-0092-7}{Fast {Ewald} summation for free-space {Stokes} potentials}, Research in the Mathematical Sciences 4~(1) (2017) 1.
\newblock \href {http://dx.doi.org/10.1186/s40687-016-0092-7} {\path{doi:10.1186/s40687-016-0092-7}}.
\newline\urlprefix\url{https://doi.org/10.1186/s40687-016-0092-7}

\bibitem{deserno_how_1998}
M.~Deserno, C.~Holm, \href{http://aip.scitation.org/doi/10.1063/1.477414}{How to mesh up {Ewald} sums. {I}. {A} theoretical and numerical comparison of various particle mesh routines}, The Journal of Chemical Physics 109~(18) (1998) 7678--7693, publisher: American Institute of Physics.
\newblock \href {http://dx.doi.org/10.1063/1.477414} {\path{doi:10.1063/1.477414}}.
\newline\urlprefix\url{http://aip.scitation.org/doi/10.1063/1.477414}

\bibitem{johnston_sinh_2005}
P.~R. Johnston, D.~Elliott, \href{https://onlinelibrary.wiley.com/doi/abs/10.1002/nme.1208}{A sinh transformation for evaluating nearly singular boundary element integrals}, International Journal for Numerical Methods in Engineering 62~(4) (2005) 564--578.
\newblock \href {http://dx.doi.org/https://doi.org/10.1002/nme.1208} {\path{doi:https://doi.org/10.1002/nme.1208}}.
\newline\urlprefix\url{https://onlinelibrary.wiley.com/doi/abs/10.1002/nme.1208}

\bibitem{liu_coupling_2004}
Y.~Liu, L.~Zhang, X.~Wang, W.~K. Liu, \href{https://onlinelibrary.wiley.com/doi/abs/10.1002/fld.798}{Coupling of {Navier}–{Stokes} equations with protein molecular dynamics and its application to hemodynamics}, International Journal for Numerical Methods in Fluids 46~(12) (2004) 1237--1252.
\newblock \href {http://dx.doi.org/10.1002/fld.798} {\path{doi:10.1002/fld.798}}.
\newline\urlprefix\url{https://onlinelibrary.wiley.com/doi/abs/10.1002/fld.798}

\bibitem{wang_numerical_2009}
T.~Wang, T.-W. Pan, Z.~W. Xing, R.~Glowinski, \href{https://link.aps.org/doi/10.1103/PhysRevE.79.041916}{Numerical simulation of rheology of red blood cell rouleaux in microchannels}, Physical Review E 79~(4) (2009) 041916, publisher: American Physical Society.
\newblock \href {http://dx.doi.org/10.1103/PhysRevE.79.041916} {\path{doi:10.1103/PhysRevE.79.041916}}.
\newline\urlprefix\url{https://link.aps.org/doi/10.1103/PhysRevE.79.041916}

\bibitem{li_computational_2014}
H.~Li, T.~Ye, K.~Y. Lam, Computational analysis of dynamic interaction of two red blood cells in a capillary, Cell Biochemistry and Biophysics 69~(3) (2014) 673--680.
\newblock \href {http://dx.doi.org/10.1007/s12013-014-9852-4} {\path{doi:10.1007/s12013-014-9852-4}}.

\bibitem{xiao_simulation_2016}
L.~Xiao, Y.~Liu, S.~Chen, B.~Fu, Simulation of {Deformation} and {Aggregation} of {Two} {Red} {Blood} {Cells} in a {Stenosed} {Microvessel} by {Dissipative} {Particle} {Dynamics}, Cell Biochemistry and Biophysics 74.
\newblock \href {http://dx.doi.org/10.1007/s12013-016-0765-2} {\path{doi:10.1007/s12013-016-0765-2}}.

\bibitem{chien_blood_1967}
S.~Chien, S.~Usami, R.~J. Dellenback, M.~I. Gregersen, \href{http://science.sciencemag.org/content/157/3790/827}{Blood {Viscosity}: {Influence} of {Erythrocyte} {Deformation}}, Science 157~(3790) (1967) 827--829.
\newblock \href {http://dx.doi.org/10.1126/science.157.3790.827} {\path{doi:10.1126/science.157.3790.827}}.
\newline\urlprefix\url{http://science.sciencemag.org/content/157/3790/827}

\bibitem{javadi_silico_2021}
E.~Javadi, Y.~Deng, G.~E. Karniadakis, S.~Jamali, \href{https://www.sciencedirect.com/science/article/pii/S0006349521004422}{In silico biophysics and hemorheology of blood hyperviscosity syndrome}, Biophysical Journal 120~(13) (2021) 2723--2733.
\newblock \href {http://dx.doi.org/10.1016/j.bpj.2021.05.013} {\path{doi:10.1016/j.bpj.2021.05.013}}.
\newline\urlprefix\url{https://www.sciencedirect.com/science/article/pii/S0006349521004422}

\bibitem{zinchenko_shear_2002}
A.~Z. Zinchenko, R.~H. Davis, Shear flow of highly concentrated emulsions of deformable drops by numerical simulations, Journal of Fluid Mechanics 455 (2002) 21--61, publisher: Cambridge University Press.

\bibitem{batchelor_stress_1970}
G.~K. Batchelor, The stress system in a suspension of force-free particles, Journal of Fluid Mechanics 41~(3) (1970) 545--570, publisher: Cambridge University Press.

\bibitem{faghih_deformation_2020}
M.~M. Faghih, M.~K. Sharp, \href{https://doi.org/10.1007/s10237-019-01208-3}{Deformation of human red blood cells in extensional flow through a hyperbolic contraction}, Biomechanics and Modeling in Mechanobiology 19~(1) (2020) 251--261.
\newblock \href {http://dx.doi.org/10.1007/s10237-019-01208-3} {\path{doi:10.1007/s10237-019-01208-3}}.
\newline\urlprefix\url{https://doi.org/10.1007/s10237-019-01208-3}

\bibitem{lu_boundary_2019}
H.~Lu, Z.~Peng, \href{https://doi.org/10.1063/1.5081057}{Boundary integral simulations of a red blood cell squeezing through a submicron slit under prescribed inlet and outlet pressures}, Physics of Fluids 31~(3) (2019) 031902.
\newblock \href {http://dx.doi.org/10.1063/1.5081057} {\path{doi:10.1063/1.5081057}}.
\newline\urlprefix\url{https://doi.org/10.1063/1.5081057}

\bibitem{zhu_dynamics_2022}
Q.~Zhu, X.~Bi, Dynamics of erythrocytes in oscillatory shear flows: effects of {S}/{V} ratio, Soft Matter 18~(5) (2022) 964--974, publisher: Royal Society of Chemistry.
\newblock \href {http://dx.doi.org/10.1039/D1SM01430G} {\path{doi:10.1039/D1SM01430G}}.

\bibitem{xu_cell-scale_2023}
Z.~Xu, C.~Chen, P.~Hao, F.~He, X.~Zhang, \href{https://www.frontiersin.org/articles/10.3389/fphys.2023.1181423}{Cell-scale hemolysis evaluation of intervenient ventricular assist device based on dissipative particle dynamics}, Frontiers in Physiology 14.
\newline\urlprefix\url{https://www.frontiersin.org/articles/10.3389/fphys.2023.1181423}

\bibitem{tran-son-tay_membrane_1987}
R.~Tran-Son-Tay, S.~P. Sutera, G.~I. Zahalak, P.~R. Rao, Membrane stress and internal pressure in a red blood cell freely suspended in a shear flow, Biophysical Journal 51~(6) (1987) 915--924.
\newblock \href {http://dx.doi.org/10.1016/S0006-3495(87)83419-7} {\path{doi:10.1016/S0006-3495(87)83419-7}}.

\bibitem{sutera_deformation_1975}
S.~P. Sutera, M.~H. Mehrjardi, \href{https://www.ncbi.nlm.nih.gov/pmc/articles/PMC1334606/}{Deformation and fragmentation of human red blood cells in turbulent shear flow.}, Biophysical Journal 15~(1) (1975) 1--10.
\newline\urlprefix\url{https://www.ncbi.nlm.nih.gov/pmc/articles/PMC1334606/}

\end{thebibliography}

\newpage

\end{document}